# Design, Fabrication, Characterization and Reliability Study of CMOS-MEMS Lorentz-Force Magnetometers

J. J. Valle[a,*], J. M. Sánchez-Chiva[a,b], D. Fernández[c] and J. Madrenas[a]

[a]Department of Electronic Engineering, Universitat Politècnica de Catalunya, Jordi Girona 1 i 3, Edifici C4, 08034 Barcelona, Spain
[b]Sorbonne Université, CNRS, Laboratoire de Recherche en Informatique (LIP6), UMR7606, 4 place Jussieu, 75005 Paris, France
[c]Institut de Física d'Altes Energies (IFAE), The Barcelona Institute of Science and Technology (BIST), Edifici Cn. Facultat Ciències Nord, Universitat Autònoma de Barcelona, 08193 Bellaterra (Barcelona), Spain

Abstract

This article presents several design techniques to fabricate micro-electro-mechanical systems (MEMS) using standard complementary metal–oxide semiconductor (CMOS) processes. They were applied to fabricate high yield CMOS-MEMS shielded Lorentz-force magnetometers (LFM). The multilayered metals and oxides of the back-end-of-line (BEOL), normally used for electronic routing, comprise the structural part of the MEMS. The most important fabrication challenges, modeling approaches and design solutions are discussed. Equations that predict the Q factor, sensitivity, Brownian noise and resonant frequency as a function of temperature, gas pressure and design parameters are presented and validated in characterization tests. A number of the fabricated magnetometers were packaged into Quad Flat No-leads (QFN) packages. We show this process can achieve yields above 95 % when the proper design techniques are adopted. Despite CMOS not being a process for MEMS manufacturing, estimated performance (sensitivity and noise level) is similar or superior to current commercial magnetometers and others built with MEMS processes. Additionally, typical offsets present in Lorentz-force magnetometers were prevented with a shielding electrode, whose efficiency is quantified. Finally, several reliability test results are presented, which demonstrate the robustness against high temperatures, magnetic fields and acceleration shocks.

## Introduction

Today, the most common form of mass-production semiconductor device fabrication is CMOS technology. The dedicated integrated circuit (IC) interfaces of commercial sensors are realized using this technology. However, the sensing elements need to be manufactured using specialized micro-machining processes. Integration of CMOS electronics and MEMS devices on a single chip (CMOS-MEMS) has the potential of reducing fabrication costs, size, parasitics and power consumption, compared to other integration approaches [1]. Remarkably, a CMOS-MEMS device may be built with the back-end-of-line (BEOL) layers of the CMOS process [2–5]. Despite its advantages, this approach has proven to be very challenging given the current lack of commercial products in the market.

In this work, we present and discuss the challenges, modeling and design techniques used to fabricate high-yield CMOS-MEMS devices. They were applied to fabricate integrated Lorentz-force magnetometers (LFM) which were packaged, characterized and subjected to several reliability tests. These are missing in most technical works in the literature, yielding the practical commercialization of LFMs still unknown.

All commercial magnetometers are non-Lorentz-force ones. They are typically based on the Hall effect, anisotropic magnetoresistance (AMR), giant magnetoresistance (GMR), magnetic tunnel junction (MTJ), or, recently, tunnel magnetoresistance (TMR) [6]. They all have some sort of magnetic material, like flux concentrators [7]. The magnetic material may be damaged by high magnetic fields, imposes temperature limitations and is susceptible to magnetic hysteresis, which in turn, may lead to reduced accuracy and require tedious re-calibration from the user. Although LFM do not require magnetic materials, they suffer from other offsets related to electrical interference [8–11]. This will be analyzed, solutions will be presented and their efficiency quantified.

### Lorentz-force magnetometers: Principle of operation and analysis

The operational principle of the shielded LFM discussed in this work is illustrated in Fig. 1. The magnetic field $\overrightarrow{B}$ is sensed with a current $\overrightarrow{i_L}$ (Lorentz current) that flows along a wire (Lorentz wire) of length $L$ inside a movable

juanvallefraga@gmail.com (J.J. Valle); jose.sanchez_chiva@sorbonne-universite.fr (J.M. Sánchez-Chiva); dfernandez@ifae.es (D. Fernández); jordi.madrenas@upc.edu (J. Madrenas)
ORCID(s): 00000-0001-9849-7868 (J.J. Valle); 0000-0002-1101-6804 (J.M. Sánchez-Chiva); 0000-0002-1076-6697 (D. Fernández); 0000-0001-5905-9179 (J. Madrenas)

structure, which experiences a force $F_m$ given by:

$$\overrightarrow{F_m} = L \cdot \overrightarrow{i_L} \times \overrightarrow{B} \tag{1}$$

The system behaves as a damped harmonic oscillator. In this work, the movable structure is formed by several ($n_B$) clamped beams, and the Lorentz current passes along each beam $n_w$ times. This increases the effective Lorentz current. Additionally, $i_L$ is applied as a square wave whose frequency is coincident with the first resonant frequency of the structure ($f_r$) in order to maximize the output or sensed current ($i_{sense}$). The Lorentz wire is shielded from the sense electrode, so $i_{sense}$ is independent of $i_L$, and this provides some important benefits that will be examined later.

In order to sustain the oscillation when $\overrightarrow{B} = 0$, or simply to characterize the device, an electrostatic driving is applied between the shield and sense electrodes. This creates an electrostatic force labeled as $F_e$ in Fig. 1, which allows to track the resonance frequency as done in many works[10,12–17].

Also, the figure shows the most important electrical parameters of the whole system.

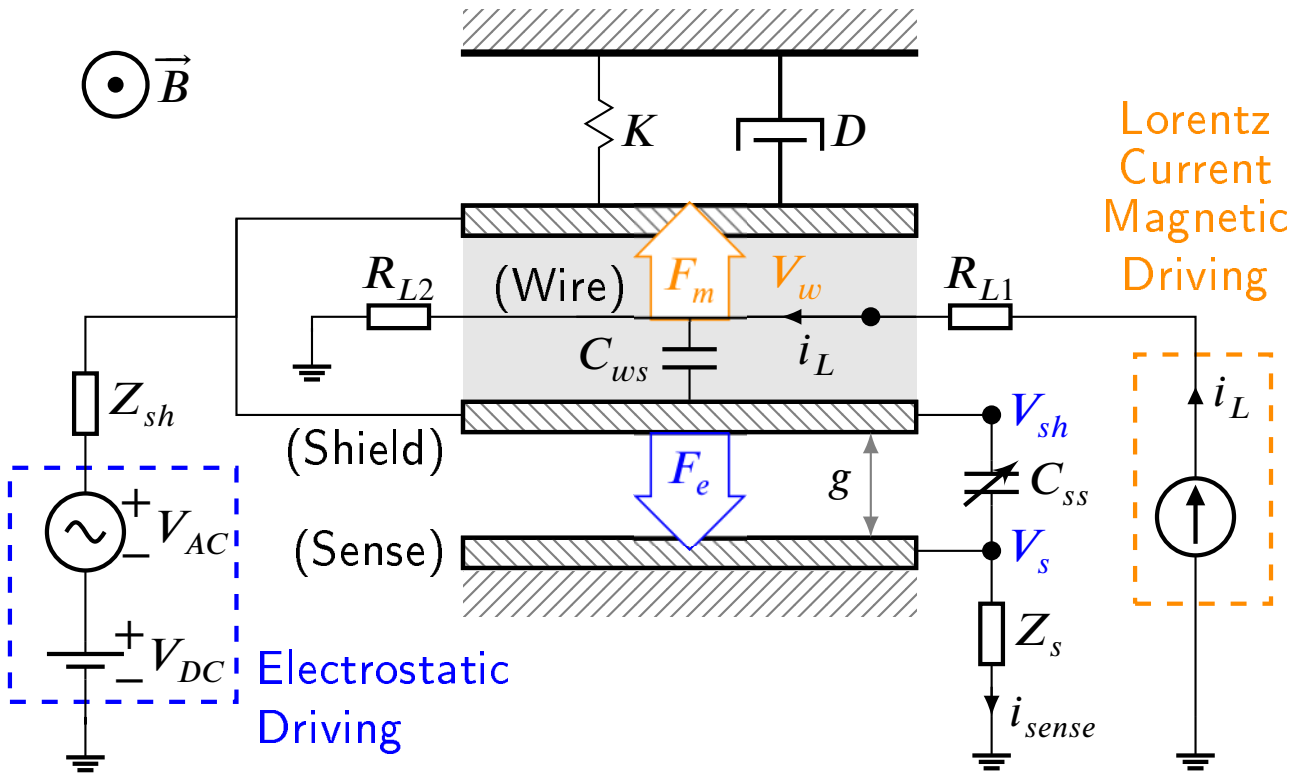

**Figure 1:** Principle of operation of a shielded Lorentz-force magnetometer.

### *Sensitivity and offsets induced by the Lorentz current*

The electrostatic force $F_e$ is a function of the gap ($g$) between plates, the sensing area ($A$) and the voltage difference ($\Delta V = V_{sh} - V_s$). For the typical impedance values and operating frequencies $V_{sh}/V_s > 10^4$, so $\Delta V \approx V_{sh}$. Additionally, the Lorentz wire AC voltage ($V_w$) induces an interference voltage ($V_{sh}^{int}$) in the shield electrode, yielding $V_{sh} \approx V_{DC} + V_{AC} + V_{sh}^{int}$, assuming small $Z_{sh}$. These considerations allow to write $i_{sense}$ and $F_e$ as:

$$i_{sense} = \frac{dQ_{ss}}{dt} = \frac{d\left(C_{ss}\Delta V\right)}{dt} \approx \underbrace{C_{ss}\frac{dV_{AC}}{dt}}_{\text{Electrical coupling}} + \underbrace{C_{ss}\frac{dV_{sh}^{int}}{dt}}_{\text{Electrical interference}} + \underbrace{V_{sh}\frac{dC_{ss}}{dt}}_{\text{Motional current } (i_m)} \tag{2}$$

where $Q_{ss}$ is the charge of $C_{ss}$. The motional current ($i_m$) is caused by the magnetic field ($B$) and its associated Lorentz force ($F_m$), and/or by the electrostatic force ($F_e$)[18]:

$$F_e = \frac{\partial}{\partial g}\left[\frac{1}{2}C_{ss}\left(V_{sh} - V_s\right)^2\right] \approx \frac{V_{sh}^2}{2}\frac{\partial C_{ss}}{\partial g} \approx \underbrace{-\frac{\epsilon_0 A}{2g^2}V_{DC}^2}_{F_{dc}(\omega=0)} \underbrace{-\frac{\epsilon_0 A}{g^2}(\overbrace{V_{DC}V_{AC}}^{\text{Electrostatic driving}} + \overbrace{V_{DC}V_{sh}^{int}}^{\text{Electrostatic interference}})}_{F_{\omega_r}(\omega=\omega_r)} \underbrace{-\frac{\epsilon_0 A}{2g^2}(V_{AC} + V_{sh}^{int})^2}_{F_{2\omega_r}(\omega=2\omega_r)} \tag{3}$$

where a simple parallel plate configuration was assumed for now. Only the $F_{\omega_r}$ terms excite the device at its resonant frequency and therefore contribute to motional current at $\omega_r$. The $V_{DC}V_{AC}$ term corresponds to the expected electrostatic driving force. The $V_{DC}V_{sh}^{int}$ term is an electrostatic interference caused by the Lorentz wire: it causes an undesired mechanical resonance (magnetic offset) that is, unfortunately, indistinguishable from the one caused by a constant external magnetic field $B$ in un-shielded LFM. The underlying reason is that $B$ and $V_{sh}^{int}$ are both proportional to the Lorentz current. Interestingly, in a shielded LFM this current is in quadrature with the magnetic field sense current. However, its suppresion with electronic techniques is challenging.

Two interference mechanisms due to the Lorentz wire have been identified: the electrical interference in Eq. (2) and the electrostatic interference in Eq. (3). Ultimately, both processes appear as an offset in the measured magnetic field. In conventional unshielded magnetometers this offset is generally orders of magnitude larger than the magnetic signal[9,19]. In fact, this is a well-known drawback of resonant Lorentz-force magnetometers[8–11]. The offset can be compensated, for example, using a DC compensation voltage applied to the MEMS structure to null the electrostatic force[8,19], but it is difficult to eliminate its associated drift reliably and cheaply[9]. There is a patented technique[10,12,20], based on current chopping, that reduces this effect greatly. It will be used in the Offset and shielding efficiency section.

The motional current in Eq. (2) depends on the applied force ($F_m + F_e$). For a damped harmonic oscillator driven at the resonant frequency:

$$i_m = V_{sh}\frac{dC_{ss}}{dt} = V_{sh}\frac{dC_{ss}}{dg}\frac{dg}{dt} = V_{sh}\frac{dC_{ss}}{dg}j\omega_r Q\frac{F_m + F_e}{K} \tag{4}$$

where $j$ indicates 90° phase and $Q$ is the quality factor of the resonator, and $K$ the spring constant.

The sensitivity $S$ of the device to the external magnetic field $B$ may now be derived by combining Eqs. (1) to (4):

$$S = \frac{\partial i_m}{\partial B} = V_{sh}\frac{dC_{ss}}{dg}j\omega_r Q\frac{i_L n_w n_B L}{K} \tag{5}$$

### *Noise and heading accuracy*

Three devices aligned along the three Cartesian axes form a three-axis magnetometer that can work as a magnetic compass. The best heading accuracy, or angle error, of a magnetic compass sensor is limited by the Brownian noise. It can be calculated as the ratio between the equivalent magnetic field noise ($B_{noise}$) and the Earth's magnetic field ($B_{earth}$):

$$\theta_{noise}^{RMS}(°/\sqrt{\text{Hz}}) = \text{atan}\left(\frac{B_{noise}^{RMS}}{B_{earth}}\right) \approx \frac{B_{noise}^{RMS}}{B_{earth}} \cdot \frac{180}{\pi} \tag{6}$$

where a linear approximation is valid given that $B_{noise} << B_{earth}$. Earth's magnetic field horizontal intensity ranges from around 40 µT in Southeast Asia to 15 − 20 µT in areas like South America, South Africa, Siberia and northern Canada[21,22]. So, regarding heading accuracy, we think that considering $B_{earth} = 20\,\mu\text{T}$ in Eq. (6) is a reasonable assumption.

To calculate $B_{noise}^{RMS}$ let us first calculate the total Lorentz force $F_m$ per unit of magnetic field $B$:

$$\frac{F_m^{RMS}}{B} = i_L^{RMS} \cdot L \cdot n_w \cdot n_B = i_L \cdot \frac{4}{\pi} \cdot \frac{1}{\sqrt{2}} \cdot L \cdot n_w \cdot n_B \tag{7}$$

where $L$ is the length of the beams and the RMS Lorentz current component at the resonant frequency ($i_L^{RMS}$) is calculated with the square wave Lorentz current $i_L$:

$$i_L^{RMS} = i_L \cdot \frac{4}{\pi} \cdot \frac{1}{\sqrt{2}} \text{ (Square wave)} \tag{8}$$

which was used to obtain the last term of Eq. (7). On the other hand, the Brownian noise force ($F_{noise}$) is given by[23]:

$$F_{noise}^{RMS} = \sqrt{4K_B T D} = \sqrt{\frac{4K_B T M\omega_r}{Q}} \tag{9}$$

where $D = M\omega_r/Q$ is the damping coefficient of the considered 1-D or lumped system of mass $M$.

Using the previous equations allows us to write the equivalent magnetic field noise in units of T/$\sqrt{\text{Hz}}$ as:

$$B_{noise}^{RMS} = \frac{F_{noise}^{RMS}}{F_{m}^{RMS}/B} = \frac{\pi\sqrt{K_B T M \omega_r}}{i_L^{peak} \cdot L \cdot n_w \cdot n_B \cdot \sqrt{2Q}} \tag{10}$$

Eqs. (6) and (10) will be evaluated for the fabricated devices later in the Results and discussion section.

# Materials and methods

## CMOS-MEMS fabrication process

The CMOS-MEMS process used in this work uses the back-end-of-line (BEOL) of a standard 6-metal 0.18 µm CMOS process to build the MEMS.The unwanted inter-metal-dielectric (IMD) oxide is etched away with a vapor HF (vHF) process already described in previous works[5,24,25]. The vHF enters through small holes in the last metal of the BEOL (detail E of Fig. 2), dissolving the oxide selectively and releasing the MEMS structure (details F and G of Fig. 2). For fabrication reasons, each BEOL layer is composed of several sub-layers[5], which makes the mechanical modelization of the final multilayered structure difficult. Finally, the MEMS devices are sealed in vacuum with a post-CMOS layer of Aluminum (detail D of Fig. 2) deposited on top of the last metal, diced and packaged[26].

We mostly used the 1P6M 0.18 µm CMOS process from Global Foundries (GF), but similar processes from LFoundry (1P6M 0.15 µm) and TSMC (1P6M 0.18 µm) worked well, also.

The fabrication process is simple, but using the BEOL as structural layers has important drawbacks such as non repeatability, excessive curvature and creep[5]. In addition, although the IMD oxide etching rate is uniform, the vHF etch is highly catalyzed along the metals[24] producing a runaway etch that must be stopped. This increases horizontal etching speed and reduces greatly etching isotropy. We refer to this as the capillarity effect.

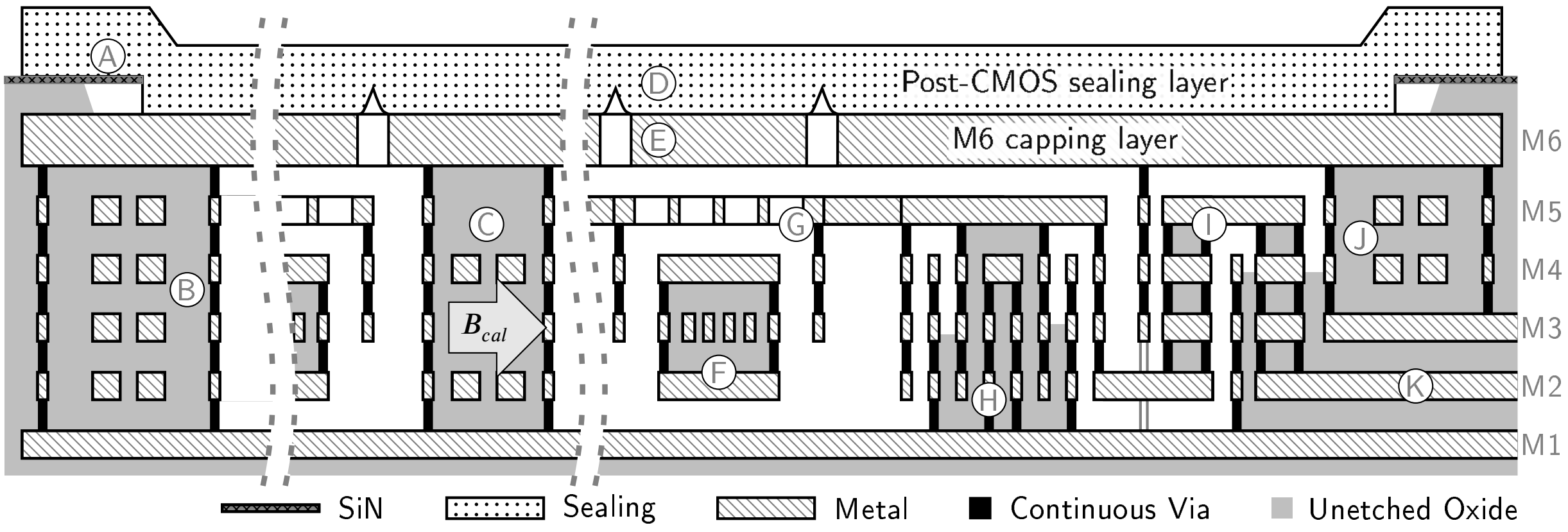

**Figure 2:** General cross-section. CMOS BEOL and post-CMOS sealing layer.

## Design techniques to overcome the CMOS-MEMS fabrication process challenges

### *Continuous vias to stop vHF*

Vias (or plugs) are generally square, tungsten-based and used for connecting two different metal levels in CMOS design. However, vias may be made very long in one direction (continuous vias) and thus fill completely the gap between two metal layers with tungsten, which is a vHF-resistant material. Vertical metal walls that stop the advance of vHF can be created this way. Although via detaching problems have been observed with other release agents[27], vHF has never caused these issues in our experience. This technique opens a myriad of possible structures for CMOS-MEMS design. One example are anchors, as the one supporting the sensing electrode in detail H of Fig. 2.

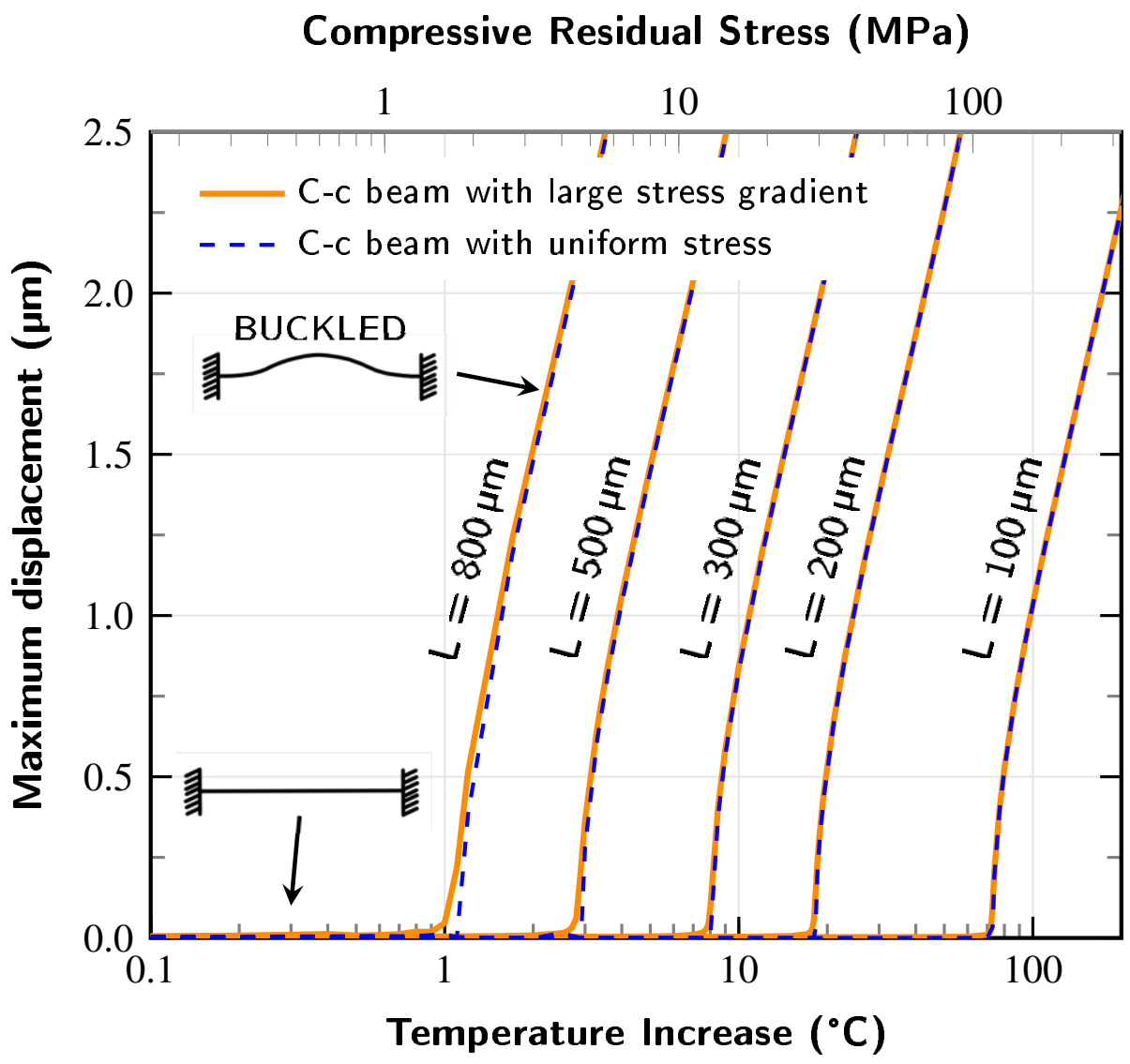

**Figure 3:** FE simulations prove the negligible effect of stress gradient on c-c beams below the critical stress or temperature, above which the beam buckles.

### *Anchors to attach mechanically and isolate electrically*

The resonating beams and driving/sensing electrodes must be mechanically attached to something and, typically, electrically isolated. Attaching them to unetched oxide proved unreliable and complicated due to the capillarity effect, which leads to very fast etching around the edges of the metals, and quick detachment from any unetched oxide. When electrical isolation is not needed, one option is to attach them to vertical walls created with metals and continuous vias. When it is needed, one compact option is to use anchors as in detail H of Fig. 2, where they provide mechanical support to the sensing electrode. They work by forcing the vHF to travel upwards and downwards, taking advantage of the slower vertical etch rate, and keeping under control the capillarity effect this way[24].

Anchors leave some oxide exposed to the MEMS cavity which may outgas if the device undergoes sufficiently high and long temperature excursions. We have found that, in devices that require a vacuum level under 1 mbar the exposed oxide should be minimized. Therefore, using as few anchors as possible may be a good design strategy in terms of outgassing minimization.

### *Clamped-clamped beams to overcome curvature issues*

Large curvature and variability are observed in CMOS-MEMS BEOL structures[3–5,25]. Fortunately, clamped-clamped (c-c) beams made of several stacked BEOL layers may be an excellent design option to circumvent these issues.

The reason is that, as long as a given critical temperature/compressive residual stress is not reached, the c-c beams will remain very flat even when there is a very large stress gradient. Simulations predict that c-c beams below the buckling load display deformations in the nanometer range (see Fig. 3) while, if only clamped at one end, they would deform from a few to hundreds of microns (as predicted by equation 7 from Valle et al.[5]). All long structures are doubly clamped in this work.

### *Several beams coupled to improve SNR and repeatability*

If $n_B$ beams are mechanically coupled they will behave as a single resonating structure. The damping coefficient $D$ is proportional to the number of coupled beams $n_B$. Therefore, the Brownian force noise (Eq. (9)) is only proportional to $\sqrt{n_B}$. By contrast, the total Lorentz force $F_m$ is proportional to $n_B$ (Eq. (7)). As a consequence, the equivalent magnetic noise (Eq. (10)) is proportional to $1/\sqrt{n_B}$, which implies that coupling more beams lowers the noise of the system. In addition, improved repeatability is also expected as variations in geometric or material properties are averaged when several beams are coupled. On the other hand, the device occupies larger space, and power consumed by the Lorentz current is increased.

### *Beam design: Offset prevention via shielding and Lorentz multiwire*

The two inherent offsets present in LFM which are caused by the Lorentz current were discussed in the introduction. In the present work both offsets and their drifts are prevented by design: firstly, the Lorentz wires are decoupled from the sensing electrodes using a metal shield around the wires as described in Sánchez-Chiva et al.[25], and as depicted in the cross-section of the beam in detail F of Fig. 2 and Figs. 4, 6a, 6b and 7. Secondly, a symmetric Lorentz wire routing design with respect to the shield electrode was adopted ($R_{L1} = R_{L2}$ in Fig. 1). This way, the central point of the wire may be kept at constant voltage: half the total voltage drop ($V_w = 0.5 \cdot V_{drop}$). Then, the interference due to the AC voltage in the rest of the wire would cancel out due to symmetry, apart from fabrication variability. Hence, little or no compensation techniques are needed. Later on, this will be shown experimentally.

Equation (11) was derived from the model in Fig. 4. It quantifies the shielding efficiency, which depends on the capacitance between the shield and the Lorentz wire ($C_{ws}$), and on the impedance between the shield and its grounding ($Z_{sh}$). As it turns out, the capacitance between the shield and the sense electrodes ($Css$) is irrelevant in practice as its associated impedance is substantially larger than $Z_{sh}$. Typical values yield $V_{sh}/V_w < 10^{-5}$.

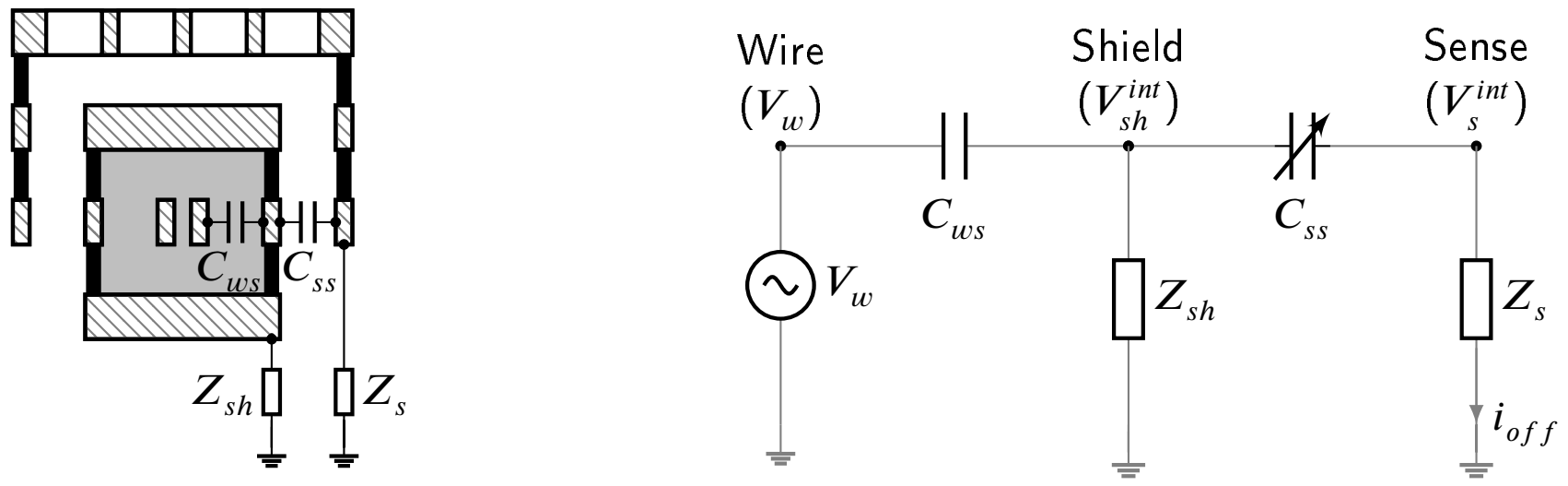

**Figure 4:** Simplified equivalent ac circuit for shielding efficiency ($V_{sh}^{int}/V_w$) calculation.

$$\frac{V_{sh}^{int}}{V_w} \approx j\omega \cdot Z_{sh} \cdot C_{ws} \tag{11}$$

The metal shield provides additional advantages: it encloses unetched oxide along with one or two metal layers that can be used to route several Lorentz wires ($n_w > 1$) along each beam (multiwire beams), as illustrated in the beam cross-sections of Figs. 6a, 6b and 7. Increasing the Lorentz current per beam delivers several obvious benefits, such as sensor area reduction and improvement of sensitivity (Eq. (5)), signal-to-noise ratio (SNR) (Eq. (10)) or heading accuracy (Eq. (6)). The number of wires per beam has limitations, though, such as electromigration current, maximum resistance allowed for the Lorentz wire due to maximum supply voltage, or filtering limitations due to the impedance of the wire at high frequency. Usually, electromigration current is the limiting factor.

### *Electrostatic sensing/actuation techniques*

A 3D magnetometer demands both vertical and horizontal sensing electrodes. Devices that vibrate vertically detect in-plane xy magnetic field. Conversely, devices that vibrate horizontally detect out-of-plane z magnetic field as Eq. (1) dictates. Fortunately, the CMOS BEOL layers allow multiple design options (see yellow electrodes of devices in Figs. 6a, 6b and 7). The cross-section of a typical lateral sensing design is depicted in the zoomed red box of Fig. 7. Two different vertical sensing designs are shown in Figs. 6a and 6b. Note that the sensing electrode in Fig. 6a has a vertical flange at each end: they act as stiffeners that reduce potential curling caused by the stress gradient. One advantage of the vertical sensing design of Fig. 6b is that it allows to use 4 layer beams and, therefore, have 2 layers dedicated for the Lorentz current-carrying wires.

While a large variation of capacitance versus displacement is generally desired to maximize sensitivity (Eq. (5)), other aspects such as damping and Q factor need to be taken into account. In this respect, the aforementioned designs may be improved by adding through-holes that reduce air squeezing and, therefore, damping. The vertical sensing electrode shown in Figs. 2 and 6a has these holes, for instance. Also, two z devices with $g = 0.35$ µm with lateral sensing electrodes solid as in Fig. 7 showed Q factors 1.5 smaller than the same device with sensing electrodes formed by layers joined with standard vias which let the air flow through.

The sensing electrodes are supported on anchors at both ends and at intermediate points. They were designed as clamped-clamped structures given that this was found to be the most mechanical reliable design.

In order to maximize the ratio capacitance variation versus static capacitance, the sensing electrodes of some devices are only at the central part of the beams, where the vibration amplitude is maximum, as in the devices of Figs. 6a and 6b. This is also beneficial in terms of Q and, therefore, sensitivity enhancement.

### *Electrical routings*

*Output from vHF area:* The sensing electrodes are in the etched area. As a consequence, connecting them to the electronics while avoiding the capillarity effect and thus, containing the vHF, proved challenging. The technique used in this work takes advantage of the vHF etching anisotropy, as the anchors previously described do: The output routing describes a vertical zigzag as detail I of Fig. 2 and bottom-right cross-section in Fig. 7 show.

*Lorentz routing:* The Lorentz wires run along all beams and each beam multiple times (detail B of Fig. 2), increasing this way, the total Lorentz force/current ratio substantially. For example, for a 6-beam 4-wire per beam device as in Fig. 6a, this ratio is increased 24 times with respect to a conventional LFM. Similar multiwire approaches were followed in some works[11,28,29]. This will achieve a very low noise floor, only surpassed when using piezoelectric amplification to increase SNR[30,31].

Each Lorentz wire has a return path along the external parts of the device (detail B of Fig. 2). The magnetic field created by the Lorentz wires is at frequency much higher than Earth's magnetic field. Thus, their associated magnetic forces, detected by adjacent beams, should be mechanically filtered. In addition, the return wires are placed symmetrically at both sides of the device in such a way that the total magnetic force on the beams is zero.

Also, as already stated previously, the Lorentz wire was designed symmetric with respect to the shield electrode as a first step to suppress its interference and associated offsets.

Running many more wires along each beam could further increase the force vs current ratio, but ohmnic resistance and available voltage impose limitations on the maximum length and minimum width of the Lorentz wires. For the case of devices with only one or few beams, electromigration is typically the bottleneck, and limits the minimum width for a given Lorentz current (around $1 - 2\,\mathrm{mA}/\mu\mathrm{m}$ in the processes we have used).

The Lorentz wire impedance imposes another important limitation: it increases quite abruptly above a given pole frequency ($\omega_{pole} = 1/(R_L C_{ws})$). The pole may fall close to the mechanical resonance frequency if the device has too many coupled beams and/or turns, or its resonance frequency is too high. For the devices considered in this work it is at least one order of magnitude above the resonance frequency, so it is not a problem.

Joule heating due to the Lorentz current is proportional to the intensity squared and, therefore, takes place at twice the resonance frequency and higher, so its effects are filtered out both mechanically and electronically.

*Autocalibration routing:* Some kind of autocalibration is generally very desirable in commercial sensors to compensate inherent sensitivity variations and/or offset drifts. We have implemented it by adding wires that run along the device in specific arrangements that can create a known magnetic field, or autocalibration magnetic field. These can be seen in the white oxide areas cut in Fig. 6a. The horizontal field that is created can be seen in detail C of Fig. 2. When it is activated the sensitivity can be measured and readjusted, which is a unique feature of the presented LFM. The coupled-beam arrangement of our devices allows to add the autocalibration routing very close to the Lorentz wires, reducing this way the power consumed by the autocalibration field.

### *M6 capping and sealing layer to protect MEMS before wafer sawing and packaging*

The packaging step in MEMS products will often be the decisive one in terms of yield. The design must be robust enough to withstand wafer sawing, manipulation and final encapsulation conditions[11]. Our magnetometers have a robust design but additional mechanical protection is given by covering the whole device with the top metal layer (M6 capping), as seen in detail E of Fig. 2. In Fig. 6a most of the top metal covering has been hidden to show the structures underneath. The top metal layer is grounded and it is part of the electrical shield.

Enhanced performance requires vacuum encapsulation. This is achieved with an outgassing step followed in less than 4 hours by a $3\,\mu\mathrm{m}$ aluminum sputtered layer (sealing layer) deposited on top of the device, which takes place at around $6.4\,\mu\mathrm{bar}$. It covers the release holes, as shown in the focused-ion-beam (FIB) cuts of Fig. 5 and detail D in Fig. 2, and provides additional mechanical robustness along with a very good vacuum level that will be partially lost after final packaging due to outgassing. It is patterned and etched using the passivation SiN (detail A in Fig. 2) as the etch barrier. Only the pads and the devices remain covered.

At this point the wafer is ready for undergoing standard packaging processes. QFN packaging was chosen, and characterization and yield measurements are later presented in this work.

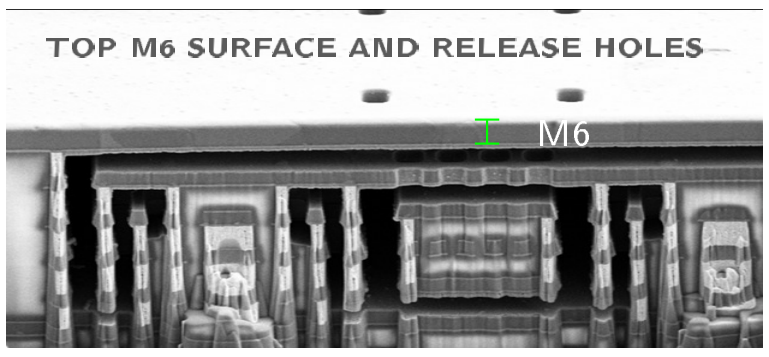

(a) XY device before sealing

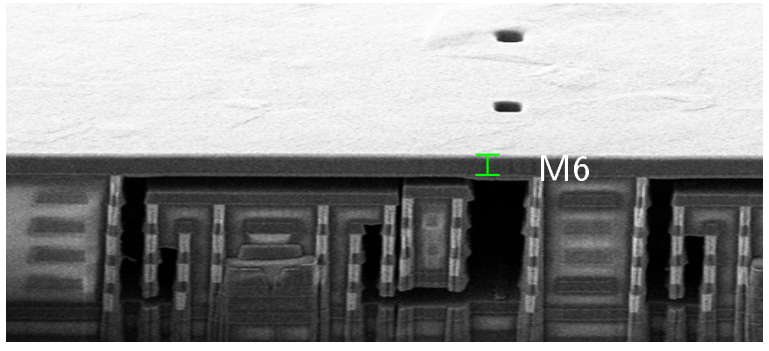

(b) Z device before sealing

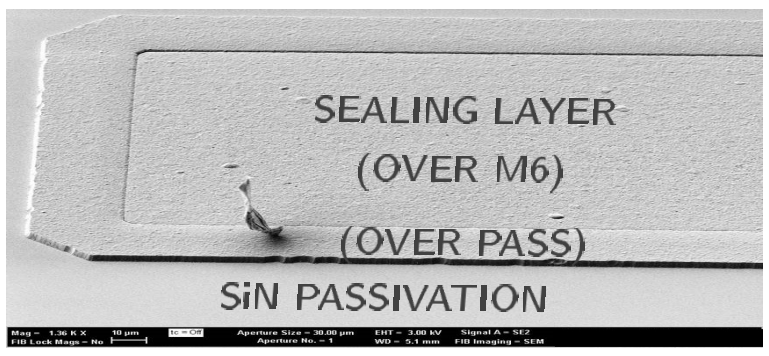

(c) Sealed device on wafer

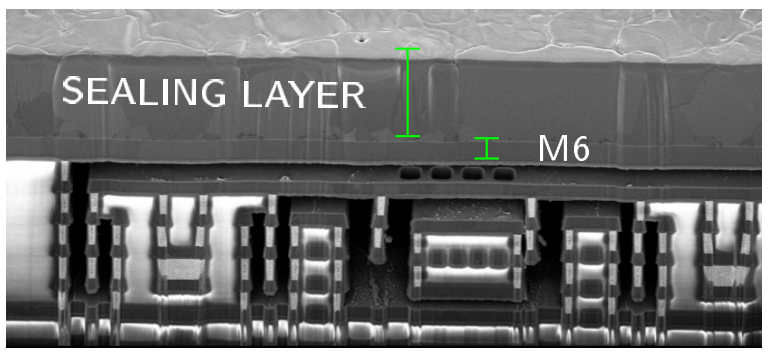

(d) XY device after 3.5 µm sealing

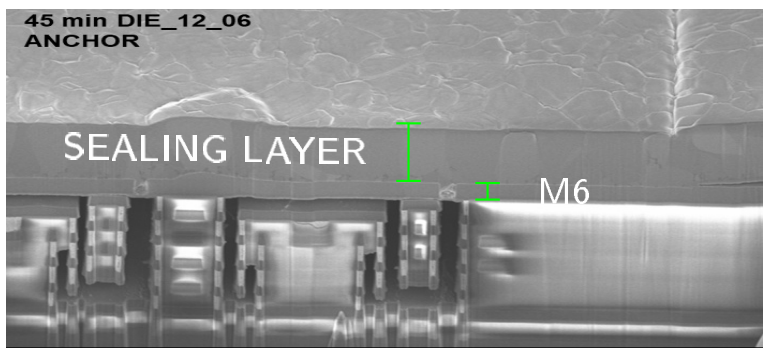

(e) Z device after 3.5 µm sealing

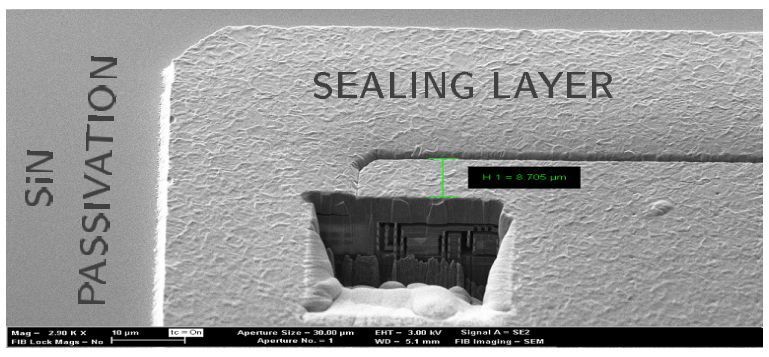

(f) Sealed device on wafer: FIB cut

**Figure 5:** Aluminum sputtering sealing result.

### *Variants of Lorentz-force CMOS-MEMS magnetometers*

The three variants analyzed in this work are shown in Figs. 6a, 6b and 7. The first two are xy devices that detect in-plane magnetic field and, therefore, resonate vertically. The third one is a z device that detects out-of-plane magnetic field and vibrates horizontally. For each type, devices with beam lengths from 80 µm to 800 µm were manufactured. This allowed to characterize and obtain reliability data as a function of the beam length.

The z device is comprised of 4-metal stack beams. The two xy devices have different sensing techniques: parallel plate in the xy device and fingers in the xy-4m device. Also, the number of stacked layers differs: 3 layers in the xy device versus 4 layers in the xy-4m device. These differences will be critical in terms of reliability as it will be demonstrated later. Most of the characterization, analysis and tests were focused on the z and xy variant.

## Modeling

In this section, the resonance frequency of clamped-clamped beams, which comprise the basic elements of the fabricated LFMs, is expressed as a function of design parameters and residual stress/temperature. Additionally, an electrical model of the MEMS, used for measuring the LFMs is developed and the most important equations derived.

### *Resonance frequency*

The resonant frequency of beams under no axial load is very well known (Blevins and Plunkett[32], Table 8-1):

$$f_0 = \frac{\lambda_i^2}{2\pi L^2}\left(\frac{EI}{m}\right)^{\frac{1}{2}} = \frac{4.73004^2}{2\pi L^2}\left(\frac{Et^2}{12\rho}\right)^{\frac{1}{2}} \tag{12}$$

where $i$ is the mode number, $E$ the Young's Modulus, $I$ the moment of inertia ($I = wt^3/12$ for a rectangular section of thickness $t$ and width $w$), m the mass per unit length, $\rho$ is the density and $\lambda = 4.73004$ for the fundamental mode ($i = 0$) and clamped-clamped conditions.

However, dealing with axially stressed clamped-clamped structures is very common in MEMS and nano design. Tensile stress ($\sigma > 0$) increases the resonant frequency and compressive stress ($\sigma < 0$) decreases it. This is a very important effect in MEMS structures that often renders Eq. (12) insufficient for correct predictions. For a given compressive load, called the critical load ($F_{cr}$) the resonant frequency is zero and the beam buckles due to elastic instability[33]. The most accurate formula that describes the frequency-residual stress dependency was given by the authors[34].

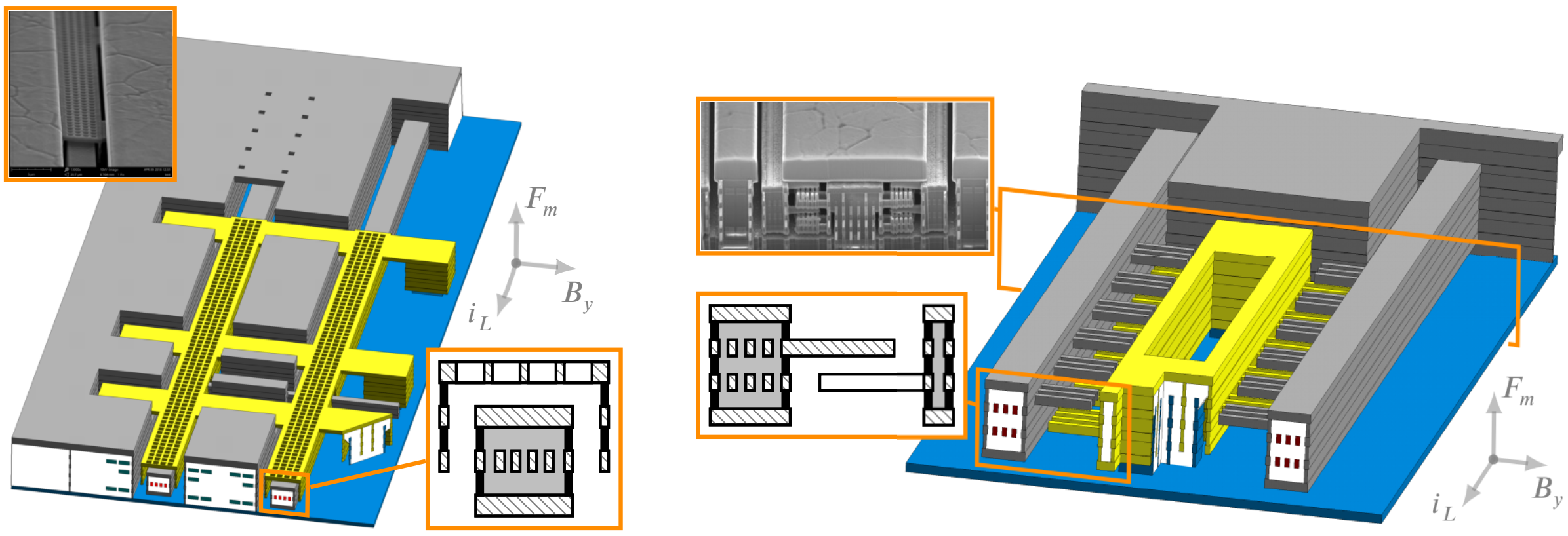

(a) XY device

(b) XY-4M device

**Figure 6:** Vertical resonance, xy devices, and cross-section

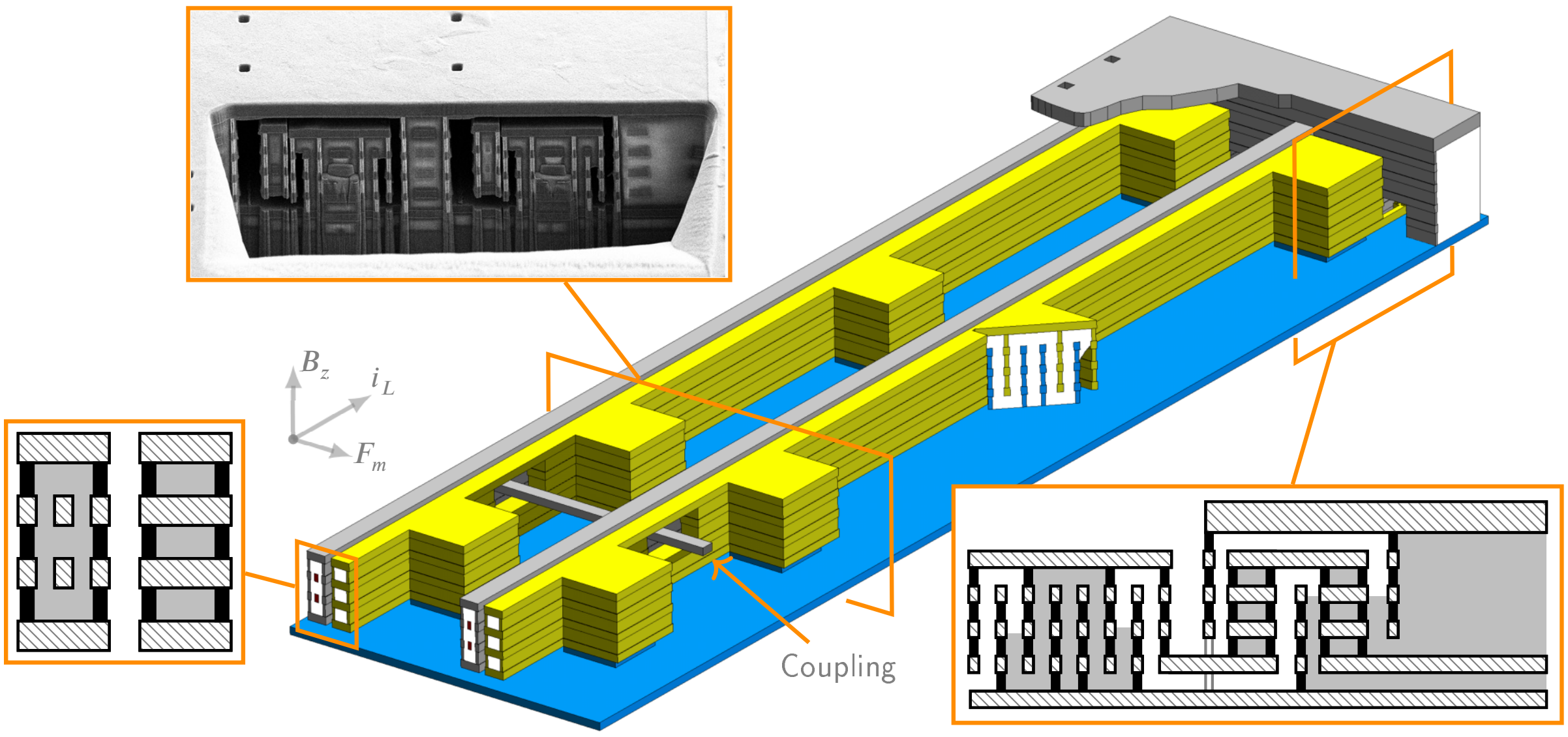

**Figure 7:** Lateral resonance, z device, and cross-section

It works in the full range of axial load ($F_a$), from the buckling point to the high-tension or string-limit regime:

$$f_r(F_a) \approx f_0 \left( 1 + \gamma \frac{F_a}{F_{cr}} + \frac{1}{\frac{1}{\alpha\gamma}\frac{F_{cr}}{F_a} + \frac{1}{\beta}} \right)^{\frac{1}{2}} \tag{13}$$

where $f_0$ is given by Eq. (12) and $\alpha = 0.19514$, $\beta = 1.2114$ and $\gamma = 0.81626$ are given in Table A.3 of Valle et al.[34] for the fundamental mode and clamped-clamped (c-c) conditions. The buckling axial load of a rectangular c-c beam is $F_{cr} = 4\pi^2 EI/L^2 = \pi^2 wt^3 E/(3L^2)$, as can be found in Table A.3 and Appendix C of Valle et al.[34]. Eq. (13) will be used to explain the observed resonant frequency of c-c beams as a function of their length and their temperature.

By using Eq. (13), the dependence with beam length can be obtained, yielding:

$$f_r \propto \frac{t}{L^2}\sqrt{\frac{E}{\rho}}\text{, when } |F_a| \ll F_{cr} \quad \text{(Small stress)} \tag{14a}$$

$$f_r \propto \frac{1}{L}\sqrt{\frac{\sigma}{\rho}}\text{, when } |F_a| \gg F_{cr} \quad \text{(Stress-dominated)} \tag{14b}$$

Note that for the beams where the tension/residual stress is the main contributor to their stiffness, the Young's Modulus $E$ or the beam thickness $t$ are of no importance for $f_r$ dependence with beam length $L$.

By assuming that the thickness $t$, the density $\rho$ and the length $L$ of the beam are constant, the resonance frequency dependence on temperature can be readily derived from Eqs. (14a) and (14b):

$$f_r \propto \sqrt{E(T)}\text{, when } |\sigma| \ll \sigma_E \quad \text{(Small stress)} \tag{15a}$$

$$f_r \propto \sqrt{\sigma(T)}\text{, when } |\sigma| \gg \sigma_E \quad \text{(Stress-dominated)} \tag{15b}$$

An implicit temperature-dependency on $E$ and the linear coefficient of thermal expansion (CTE) is contained in Eq. (15b): the axial stress $\sigma$ of a clamped-clamped beam depends on temperature due to the thermal expansion/contraction as:

$$\sigma(T) = \sigma(T_0) - \int_{T_0}^{T} \sum_i E_i \left(\alpha_i - \alpha_{subs}\right) \frac{A_i}{A}\, dT \approx \sigma(T_0) - \sum_i E_i \left(\alpha_i - \alpha_{subs}\right) \frac{A_i}{A} \left(T - T_0\right) \tag{16}$$

where sub-index $i$ refers to the material number, $\alpha$ is the beam CTE, $\alpha_{subs}$ is the substrate CTE, $A_i/A$ is the fraction of the beam section occupied by material $i$, and $\sigma(T_0)$ is the stress at an arbitrary reference temperature $T_0$. Note that the substrate also expands/contracts and needs to be taken into account in the $f_r(T)$ calculation. Typically, the Young Modulus decreases with temperature and the CTE increases with temperature. Interestingly, their product $E \cdot \alpha$ for the CMOS BEOL materials remains approximately constant, so the integral in Eq. (16) may be substituted by the temperature increment $(T - T_0)$.

### *MEMS electrical model*

The equation of motion of a driven damped spring-mass system is:

$$M\frac{\partial^2 x}{\partial t^2} + D\frac{\partial x}{\partial t} + Kx = f(t) \tag{17}$$

where $M$, $D$ and $K$ are the mass, damping coefficient and stiffness of the system, respectively, and $f(t)$ is the applied lumped force.

Let's now assume we apply a voltage $V = V_{DC} + V_{AC}$ to a movable capacitor of capacitance $C$ that is part of a damped spring-mass system, as in Fig. 1. The current flow through this moving capacitor is:

$$i = \frac{\partial CV}{\partial t} = C\frac{\partial V}{\partial t} + V\frac{\partial C}{\partial t} = C\frac{\partial V_{AC}}{\partial t} + \eta\frac{\partial x}{\partial t} \tag{18}$$

where $\eta = V_{DC}\frac{\partial C}{\partial x}$ is the electromechanical coupling factor.

The component of the electrostatic force between the plates of the capacitor at the frequency of $V_{AC}$ is:

$$F_e = \frac{\partial U}{\partial x} = \frac{\partial}{\partial x}\left(\frac{1}{2}CV^2\right) = V_{AC}V_{DC}\frac{\partial C}{\partial x} = \eta V_{AC} \tag{19}$$

It is important to note that $x$ refers to the lumped displacement used in Eq. (17), which is 0.542 times the central displacement $x_c$ of the real clamped-clamped beam. This value was calculated so that $\frac{\partial C}{\partial x} = \frac{\partial C_r}{\partial x_c}$, where $C_r$ and $C$ are the real and lumped capacitance values, respectively. This way, the lumped force $f(t)$ is equivalent to the total uniform load acting on the real beam. This is convenient for the considered devices given that the electrostatic and the Lorentz forces are applied uniformly along the beam span.

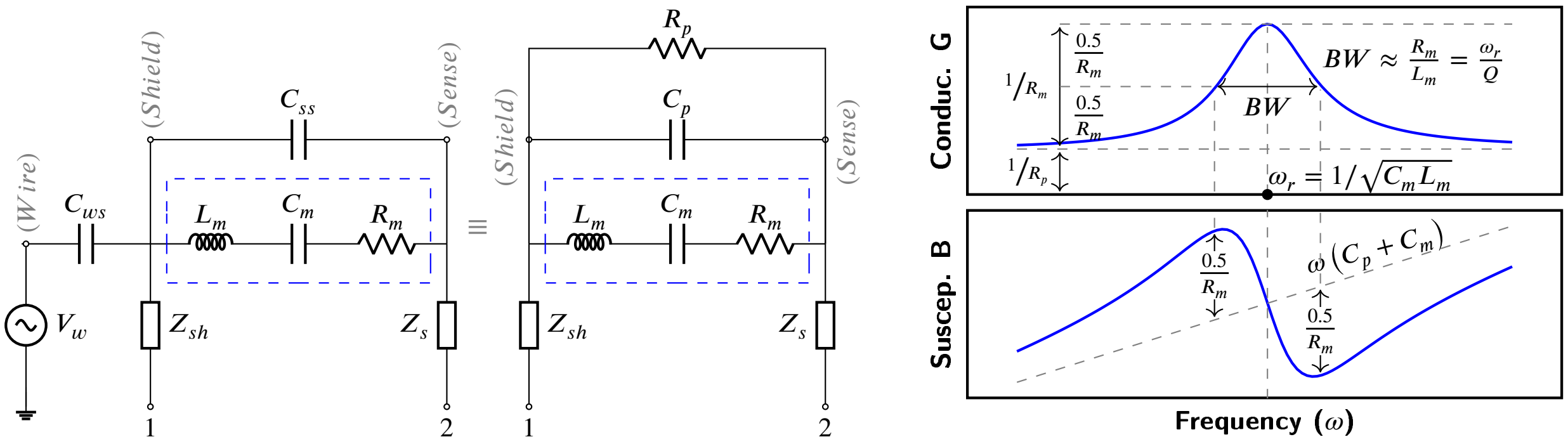

(a) Interference is included explicitly. $C_{ss}$ is the shield-sense capacitance. $C_{ws}$ is the wire-shield coupling capacitance.

(b) Lorentz current interference effects are built in $R_p$ and $C_p$.

(c) G-B curves.

**Figure 8**: Electrical equivalents of a MEMS resonator with Lorentz wire interference included (a, b). G-B example curves (c, d)

Now, defining $i_{mot} = \eta \frac{\partial x}{\partial t}$ and substituting into Eq. (17) yields:

$$\frac{M}{\eta}\frac{\partial i_{mot}}{\partial t} + \frac{D}{\eta} i_{mot} + \frac{K}{\eta}\int i_{mot} dx = f(t) \tag{20}$$

Let us assume that the force $f(t)$ is the combination of the electrostatic force $F_e$ produced by the measurement signal of an impedance analyzer and a magnetic force $F_m$. If we define the ratio $\Omega = F_m/F_e$, by substitution of $f(t)$ into Eq. (20) it is straightforward to arrive at:

$$\frac{M}{\eta^2(1+\Omega)}\frac{\partial i_{mot}}{\partial t} + \frac{D}{\eta^2(1+\Omega)} i_{mot} + \frac{K}{\eta^2(1+\Omega)}\int i_{mot} dx = L_m \frac{\partial i_{mot}}{\partial t} + R_m i_{mot} + \frac{1}{C_m}\int i_{mot} dx = V_{AC} \tag{21}$$

which corresponds to an inductor $L_m$, capacitor $C_m$ and resistor $R_m$ in series. They represent the motional inductance, capacitance and resistance values of the MEMS, respectively, and they simulate the mechanical dynamics of the MEMS resonator. It is the dashed region of the MEMS electrical model[35] used to fit the measurements in this work (see Figs. 8a and 8b).

Components $R_p$ and $C_p$ in Fig. 8b generally represent the physical electrical resistance and capacitance between the shield and sense electrodes. However, under some circumstances, the conductance $G$ and/or susceptance $B$ seen from nodes 1-2 may show calibration/interference offsets that are absorbed by $R_p$ and $C_p$, respectively. In this case, they no longer represent the physical resistance and capacitance of the MEMS. In practice these offsets are unavoidable when there is capacitive coupling between the Lorentz current wire and the sensing electrodes, as described in this and many other works[10,12,20,25]. In this work, the Lorentz wire is coupled only to the shield electrode, as shown in the explicit electromechanical model of Fig. 8a. This creates an interference signal created by $V_w$ which changes the admittance of the system seen from 1-2. Fortunately, the model in Fig. 8a can be simplified to the model in Fig. 8b, which is the one used in this work to fit G-B curves of MEMS magnetometers with and without interferences. When no current flows through the Lorentz wire there is no interference and $C_{ss} = C_p$ and $R_p \to \infty$. Finally, the impedance components $Z_{sh}$ and $Z_s$ represent the output impedance of the measurement instrument connected to 1-2.

The motional parameters are related to the mechanical and electrical properties of the system and also to $\Omega$. This is described by the following equations, derived from Eq. (21):

$$R_m = \frac{D}{\eta^2}\frac{1}{1+\Omega} = \frac{\sqrt{KM}}{Q}\frac{1}{\eta^2}\frac{1}{1+\Omega} \tag{22}$$

$$L_m = \frac{M}{\eta^2}\frac{1}{1+\Omega} \tag{23}$$

$$C_m = \frac{\eta^2}{K}(1+\Omega) \tag{24}$$

where $Q$ is the quality factor of the system:

$$Q = \frac{1}{R_m}\sqrt{\frac{L_m}{C_m}} \tag{25}$$

Note that the $L_m$, $C_m$ and $R_m$ values seen by an impedance analyzer depend on the ratio $\Omega = F_m/F_e$. When the MEMS is just characterized with an impedance analyzer, and no Lorentz current is injected $F_m = 0$ and so $\Omega = 0$. In some experiments in this work $F_m \neq 0$, like when extracting the sensitivity of the MEMS to magnetic fields.

In addition, the magnetometer sensitivity (Eq. (5)) may be rewritten in a much simpler form as a function of the parameters measured directly with the IA in units of Amperes per Tesla (A/T):

$$S = \frac{\partial i_m}{\partial B} = \frac{\Omega V_{AC}}{R_m B} \tag{26}$$

which can be further generalized in units of Amperes per Tesla and per DC voltage and Lorentz current used (A/(T V A)), as Eq. (5) shows:

$$S' = \frac{S}{V_{DC} i_L} = \frac{\Omega V_{AC}}{R_m B V_{DC} i_L} \tag{27}$$

The admittance ($Y = G + jB$) of the circuit in Fig. 8b is:

$$G(\omega) = \frac{1}{R_p} + \frac{R_m}{R_m^2 + \left(wL_m - \frac{1}{wC_m}\right)^2} \tag{28}$$

$$B(\omega) = wC_p - \frac{wL_m - \frac{1}{wC_m}}{R_m^2 + \left(wL_m - \frac{1}{wC_m}\right)^2} \tag{29}$$

where $\omega$ is the angular frequency. The G-B curves are plotted in Fig. 8c along with the circuit parameters that determine their shape. Finally, and for completeness, the lumped vibration amplitude $X$ may be written as:

$$X(\omega) = \frac{f(\omega)}{\omega\eta^2}\left(G - \frac{1}{R_p} + j(B - \omega C_p)\right) \tag{30}$$

## Methodology and measurement setup

The status of the CMOS-MEMS devices was assessed by measuring their admittance vs frequency curve $Y(\omega)$, also called G-B curve, that typically showed a resonance peak. Then, the electrical equivalent of Fig. 8b was fitted to the obtained curve. This allows accurate extraction of important electrical and mechanical parameters of the beams, like their resonant frequency and residual stress, or the capacitance between the beam and the excitation electrode, from where the status of the beam can be inferred. For example, a deformed beam which is touching the adjacent metal electrode would lead to higher conductance and no resonance peak at the expected frequency.

The G-B curve measurement was carried out with impedance analyzers (HP 4294A or Agilent E4990A) set to measure admittance values. Radio-frequency (RF) probes were used for a reduced measurement noise. A Cascade probe station 12000b was used for wafers or single dice (Fig. 9a). A socket was used for packaged samples (Fig. 9b). The beams were excited with a sinusoidal test voltage ($V_{AC}$) superimposed to a DC bias voltage ($V_{DC}$), which results in an excitation force at the frequency of $V_{AC}$ (as in Eq. (3)). The test voltages were applied between the beams and an electrode (used for both driving and sensing) which was placed either on top of the beam for exciting vertical out-of-plane vibration, or to one side for horizontal in-plane vibration excitation. In order to achieve the cleanest

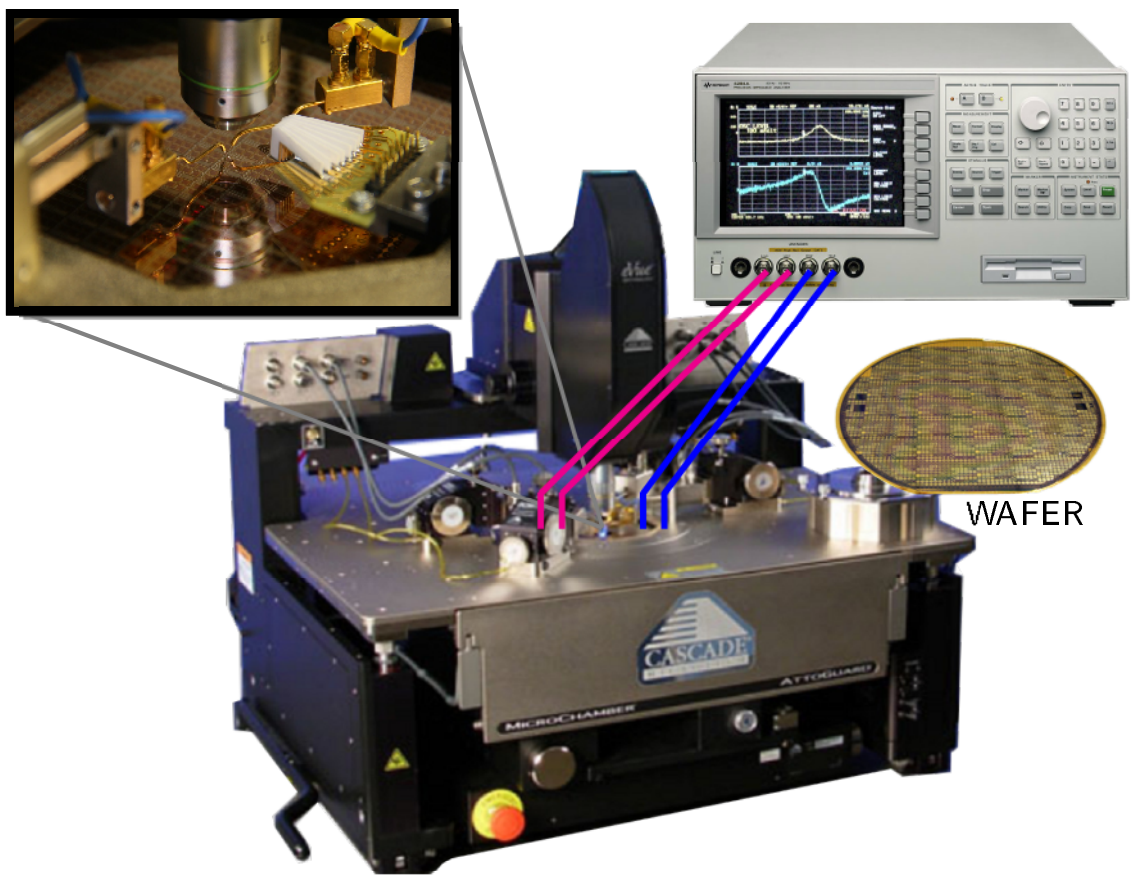

(a) On-wafer: Cascade probe station 12000b and impedance analyzer HP 4294A.

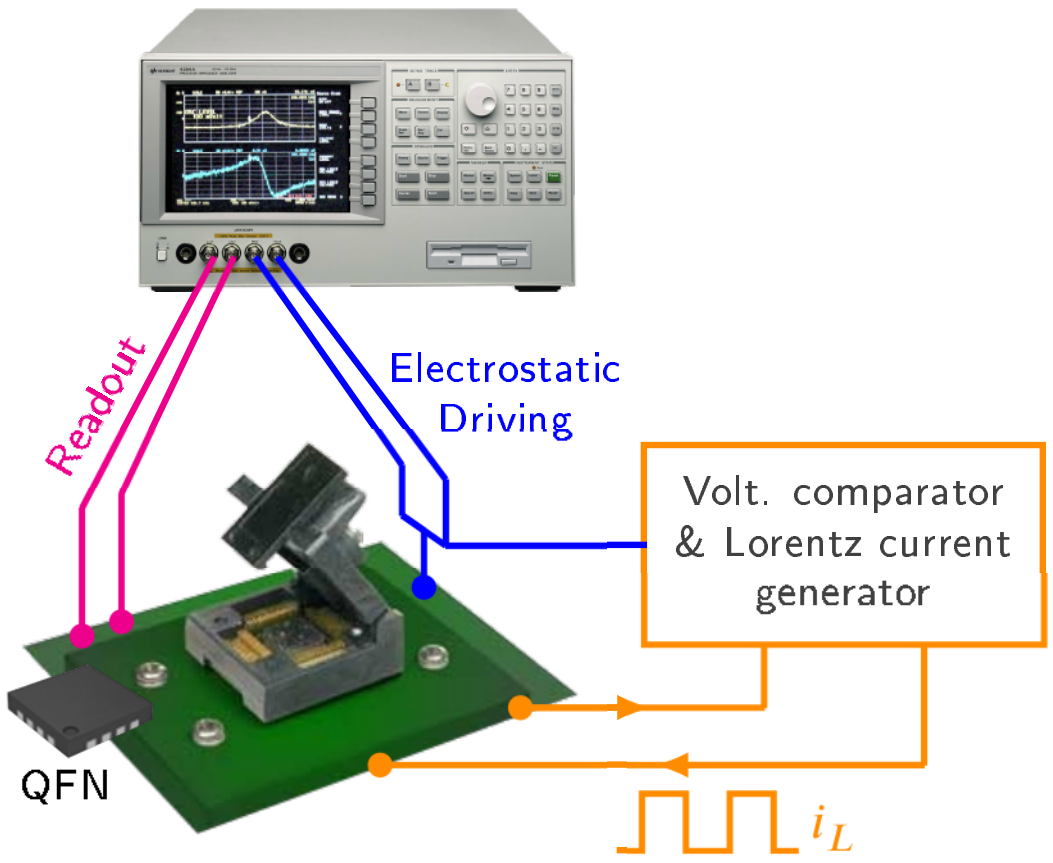

(b) Socket for QFN measurement, impedance analyzer and optional Lorentz current generation (sensitivity, offset and magnetic robustness tests).

**Figure 9:** Measurement setups

resonance curve, the test voltages were adjusted experimentally depending on a number of factors, such as quality factor, resonance frequency, parasitic capacitance, noise level and non-linear behavior of the system. A Lorentz current in phase with the AC voltage of the impedance analyzer was injected into the MEMS in some tests, namely sensitivity, offset and magnetic robustness tests. This was accomplished with the system described in Sánchez-Chiva et al.[36], for which additional equations were derived in the MEMS electrical model section: the Ω parameter plays a key role in all the tests with injected Lorentz current while the IA measures the device.

# Results and discussion

## Characterization

### *Q factor versus pressure, resonance frequency/beam length:* $Q(P, fr)$

On-wafer quality factor (Q) versus pressure (P) measurements were performed in Nitrogen (N) atmosphere at 25 °C. The results are shown in Fig. 10. They show that Q is higher the lower the pressure, reaching a saturation plateau at a pressure level that is device-dependent. This behavior obeys to the coexistence of two main damping mechanisms: air damping and intrinsic damping.

*Air Damping:* Approximately over a few mbar air damping is the main damping contributor. The Q-P curve follows the expected shape for an air damped resonator[37,38]. When the characteristic length of the structure ($L_c$) is larger than the mean free path length of the gas molecules ($\lambda_g$) the air can be modeled as a continuous viscous fluid. This is known as the fluidic regime and the Navier–Stokes equations with non-slip boundary conditions lead to[39]:

$$Q_{air} = \gamma \frac{f_r}{\mu} \quad \text{(Air damping)} \tag{31}$$

where $\gamma$ is a proportionality parameter that depends on the considered geometry and $f_r$ is the resonance frequency. The Q dependency with pressure can be introduced using a pressure-dependent artificial viscosity ($\mu$), which has been studied for different cases (squeezed-film or shear flow, molecular or slip-flow regime, diffuse or specular gas particle reflections...)[40–44]. All these approximations have a common form, which is:

$$\mu = \frac{\mu_0}{1 + \beta K_n^m} \quad \text{(Air damping)} \tag{32}$$

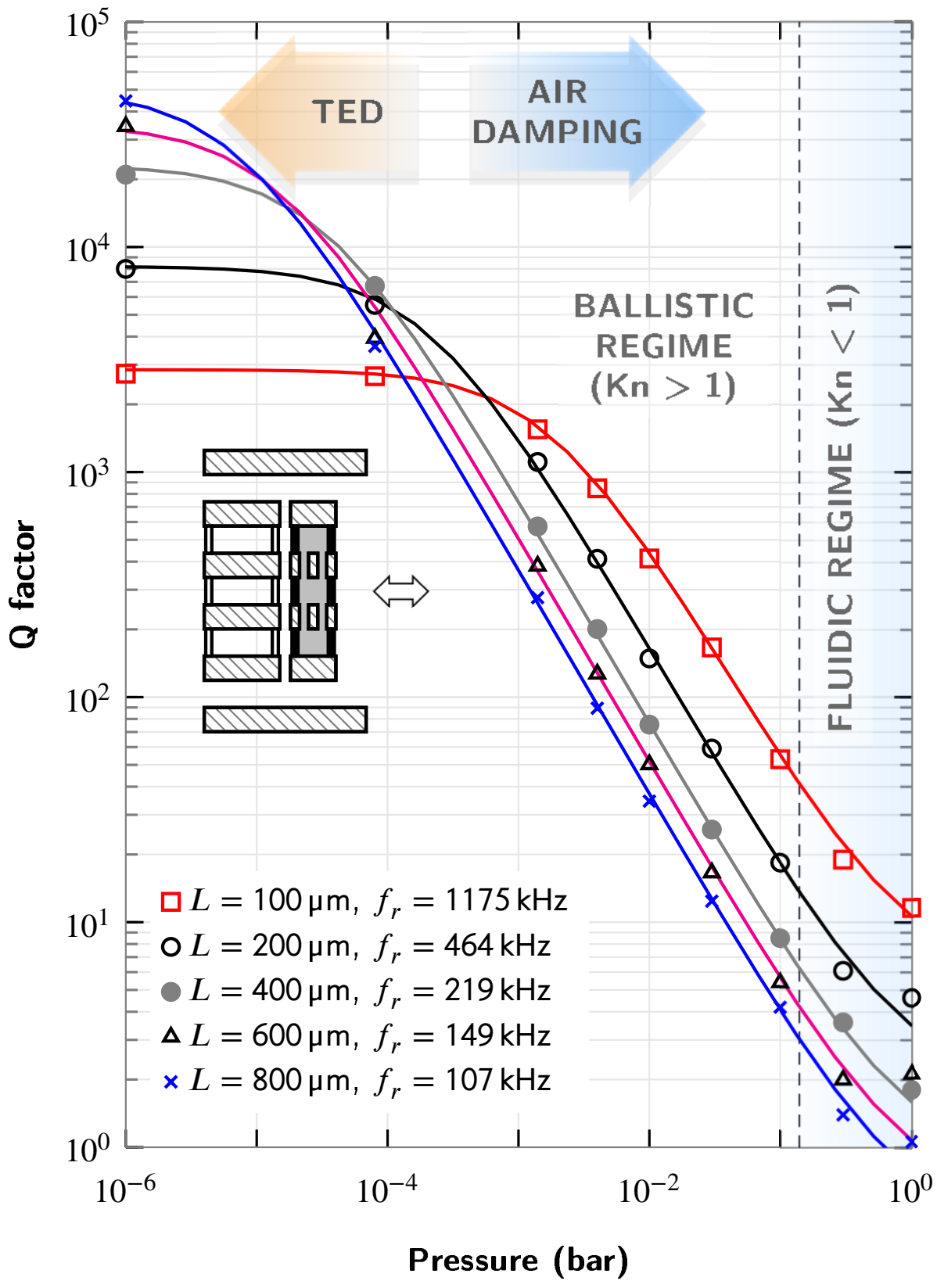

(a) Z device: Beam width = 1.5 µm, beam height = 5 µm, lateral gap = 0.35 µm.

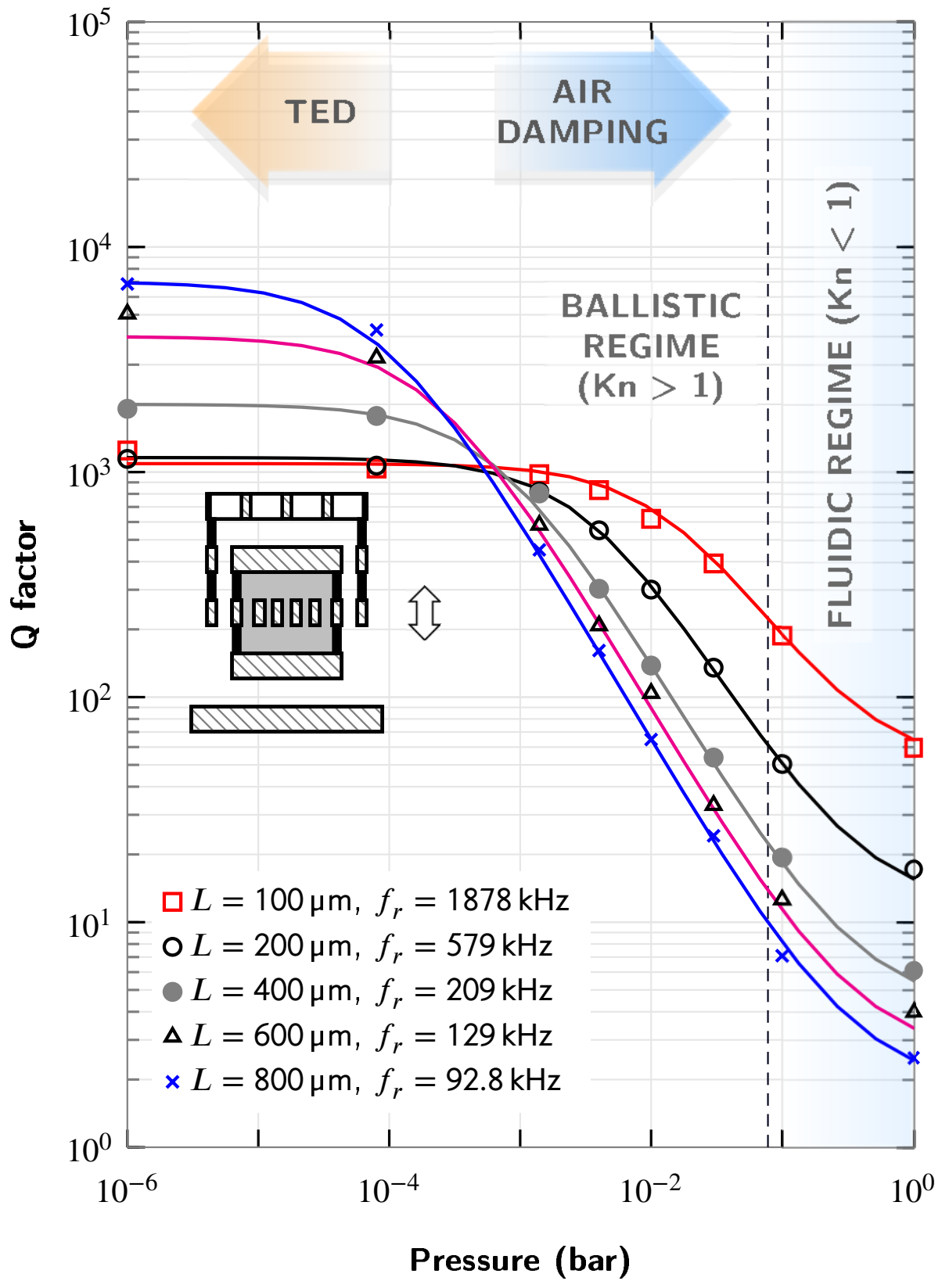

(b) XY device: Beam width = 3 µm, beam height = 3.53 µm, vertical gap = 0.9 µm, lateral gap = 0.5 µm.

**Figure 10:** Measured Q versus length or $f_r$ and pressure. The continuous lines where obtained with Eq. (38).

where $\beta$ and $m$ are two free parameters, $\mu_0$ is the dynamic viscosity of the gas at a specified temperature ($1.81 \times 10^{-5}$ Pa s at 300 K and ambient pressure), and $K_n$ is the Knudsen number:

$$K_n = \frac{\lambda_g}{L_c} \propto \frac{T}{P} \quad \begin{cases} \lambda_g^N \sim 64\,\text{nm at } P = 1atm, T = 298K \\ \lambda_g^{Air} \sim 72\,\text{nm at } P = 1atm, T = 298K \\ L_c\text{: Characteristic length} \sim \text{air gap} \end{cases} \tag{33}$$

According to the Kinetic theory of gases $\lambda_g$ is proportional to $T/P$. Remarkably, the artificial viscosity approach works reasonably well even when $K_n > 1$ and therefore the dissipation is caused not by viscous forces but by the impact of noninteracting gas molecules. This is called the ballistic or free molecular flow regime.

*Intrinsic Damping:* Intrinsic damping generally represents the Q factor upper limit at sufficiently low pressure. It arises from relaxation loss mechanisms within the resonating structure itself[38]. The better known example may be thermoelastic damping (TED), which is an absolute lower bound on intrinsic damping, but friction loss mechanisms like surface loss or phase boundary slipping in multilayer structures should also be considered in CMOS-MEMS structures. Friction loss mechanisms are a ubiquitous phenomenon and, along with TED, are best described by the Zener's anelastic relaxation theory[45,46]. In this theory, the Q factor resultant from $n$ intrinsic damping mechanisms would be given by:

$$Q^{-1}_{intrinsic} = \sum_{i=1}^{n} \Delta_i \frac{f_r/f_{Di}}{1 + (f_r/f_{Di})^2} \tag{34}$$

(2021)

| | $L_c$(nm) | $\gamma$ | $f_c$(MHz) |
|---|---|---|---|
| Z devices | 350 | $4.4\times10^{-11}$ | $\approx 1.6$ |
| Z devices* | 350 | $6.8\times10^{-11}$ | $\approx 2.3$ |
| Z devices | 500 | $8.3\times10^{-11}$ | $\approx 2.0$ |
| Z devices | 1000 | $26\times10^{-11}$ | $\approx 2.0$ |
| XY devices** | 900 | $35\times10^{-11}$ | $\approx 3.0$ |

*Sensing electrode is hollow between layers and air can flow through. **Double vertical gap of 900 nm, one of them with holes and air can flow through.

**Table 1**
Parameters used in Eq. (37).

where $\Delta_i$ is the relaxation strength and $f_{Di} = 1/(2\pi\tau_i)$ is the Debye frequency associated with the relaxation time of the ith mechanism ($\tau_i$). Generally, one mechanism is dominant and it is sufficient to consider $n = 1$. The minimum Q factor occurs when the vibration is at the Debye frequency of the dominant one. Depending on whether $f_r$ is well below $f_D$ (isothermal regime) or well above it (adiabatic regime) the Q dependency with $f_r$ is the opposite:

$$Q_{intrinsic} = \begin{cases} (f_r/f_D)/\Delta & \text{if } f_r \gg f_D \text{ (Adiabatic)} \\ 2/\Delta & \text{if } f_r = f_D \text{ (Debye Peak)} \\ (f_D/f_r)/\Delta & \text{if } f_r \ll f_D \text{ (Isothermal)} \end{cases} \tag{35}$$

In the case of TED and for uniform beams[38]:

$$\Delta_{TED} = \frac{E\alpha^2 T}{\rho C_p} \quad \text{and} \quad f_D^{TED} = \frac{\pi^2 k}{t^2 \rho C p} \tag{36}$$

where $E$ is the Young's Modulus, $\alpha$ the thermal expansion coefficient, $\rho$ the density, $C_p$ the specific heat, $k$ the thermal conduction coefficient and $t$ the beam thickness.

*Q factor characterization:* Our data clearly shows a $Q \propto P^{-1}$ dependency in the ballistic regime, which implies $m = 1$ in Eq. (32), close to most formulas in Veijola et al.[42], Li and Hughes[44]. Most authors use Veijola's formula with $m = 1.159$, intended for diffusely rejecting identical surfaces[42], but it does not work well in our case ($3 - 5$ µm wide and $100 - 800$ µm long BEOL CMOS beams with $0.35 - 1.00$ µm gaps where both slide and squeeze film damping take place).

The $\beta$ value models $Q(P)$ in the fluidic regime ($K_n \leq 1$). Typically, it may range from $\beta = 2$ for shear flow[41] to values not usually higher than 10, as shown in Li and Hughes[44]. In our case, $\beta = 5$ worked reasonably well. The proportionality factor between Q and $f_r/\mu$ in Eq. (31) defines the slope of the curve in the ballistic regime. It turns out to be $\gamma = 6.80\times10^{-11}$ Pa s$^2$ for the z devices and $\gamma = 3.50\times10^{-11}$ Pa s$^2$ for the vertical devices.

The shortest devices showed Q factors up to 30 % higher than initially expected in the air-damped region, according to their resonance frequency and Eq. (31). We think it may be caused by the air not being able to escape from the closing gap fast enough and starting to behave more like a spring and less like a damper. In this case, the damping coefficient will change with frequency $f_r$ as $\propto 1/(1 + f_r^2/f_c^2)$, where $f_c$ is the cut-off frequency[47]. The approximate cut-off frequency that best fitted the data was $f_c \approx 2.3$ MHz for the z device and $f_c \approx 3.0$ MHz for the vertical device.

After adding all the discussed corrections to Eq. (31), the Q factor due to air damping is, finally:

$$Q_{air} = \gamma \frac{f_r}{\mu_0}\left(1 + \frac{5\lambda_g}{L_c}\right)\left(1 + \frac{f_r^2}{f_c^2}\right) \tag{37}$$

where the pressure dependence is contained in $\lambda_g \propto T/P$ and $\mu_0 = f(T)$ may be considered independent of $P$. Table 1 summarizes the parameters used for the 2 cases represented in Fig. 10 (highlighted), and also provides experimental data for three additional cases.

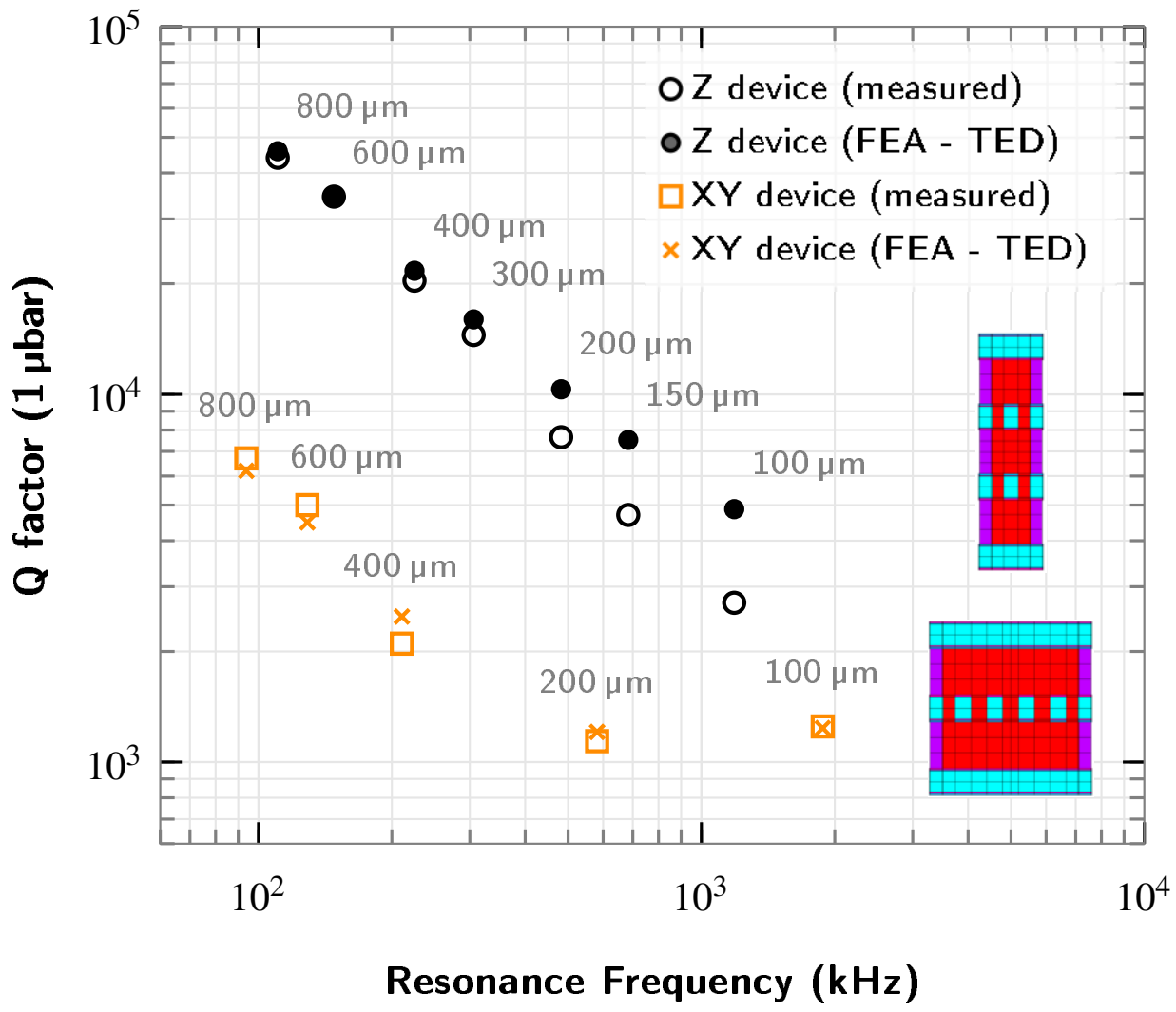

**Figure 11:** Measured Q versus frequency at $P = 1$ µbar and predicted Q factor with finite element analysis (FEA) caused by thermoelastic damping (TED). Meshed cross-sections of the two types of devices also shown.

At low pressures, another damping mechanism becomes the dominant one and the measured quality factors reach a plateau (see Fig. 10). We have plotted the measured Q factor as a function of the resonant frequency at 1 µbar in Fig. 11 in order to analyze the dominant damping mechanism. The z devices (see plateaus in Fig. 10a and circular data points in Fig. 11) operate in the isothermal region, where $Q \propto 1/f_r$, just the inverse proportionality of that found in the air-damped region. On the other hand, the shorter xy devices (Fig. 10b and square data points around 1MHz in Fig. 11) seem to operate close to their Debye frequency given than the Q factor does not depend that much on the vibration frequency.

Duwel et al.[48] has shown that TED is an important loss mechanism for flexural modes. However, Prabhakar and Vengallatorer observed in Prabhakar and Vengallatore[49] that internal friction is much higher than TED when $f_r < 1$ MHz in some bilayer structures. Given that CMOS-MEMS devices are multilayered and more complex than theirs, intrinsic friction losses might be important. However, finite element analysis (FEA) carried out by us predicted Q factors due to TED very similar to the measured ones for both types of devices. The simulated TED Debye frequency for the vertical device (0.8 MHz) is substantially smaller than for the lateral one (larger than 2 MHz). This explains the higher Q factors for the longest devices at low pressure. However, all the CMOS BEOL layers must be included in the FEA model (see cross-sections in Fig. 11), even the adhesion and antireflective coatings (Titanium and Titanium Nitride) in order to perform sufficiently accurate predictions. These layers play an important role because they have a low thermal conductance which decreases the associated TED Debye frequency and this determines greatly the simulated Q factor. Also, stress in the beams was included in the simulations given its importance in highly stressed structures[38,50]. The only deviation from simulations takes place in the shortest lateral devices, for which the TED Q factor is overestimated. It might be evidence of another damping mechanism that we have not identified.

The total Q factor is, therefore:

$$\frac{1}{Q_{total}} = \frac{1}{Q_{air}} + \frac{1}{Q_{TED}} \tag{38}$$

All the solid lines in Fig. 10 were generated with Eq. (38). The value used for $Q_{TED}$ is shown in Table 2 and was deduced from the measured data, rather than the simulation because, as already mentioned, there is some disagreement for the shortest lateral devices at 1 µbar. With that exception, Eq. (38) fits very well the measured data.

***Resonance frequency versus length:*** $f_r(L)$

The resonance frequency of magnetometers built using clamped-clamped beams of different lengths is plotted in Fig. 12. It was done for two types of beam sections and for out-of-plane and in-plane vibrations. The obtained data

(2021)

| L = | 100 µm | 200 µm | 400 µm | 800 µm |
|---|---|---|---|---|
| Z devices $Q_{TED}$ = | 2850 | 8200 | 23000 | 50000 |
| XY devices $Q_{TED}$ = | 1090 | 1160 | 2000 | 7000 |

**Table 2**
Quality factors due to TED ($Q_{TED}$) used in Eq. (38).

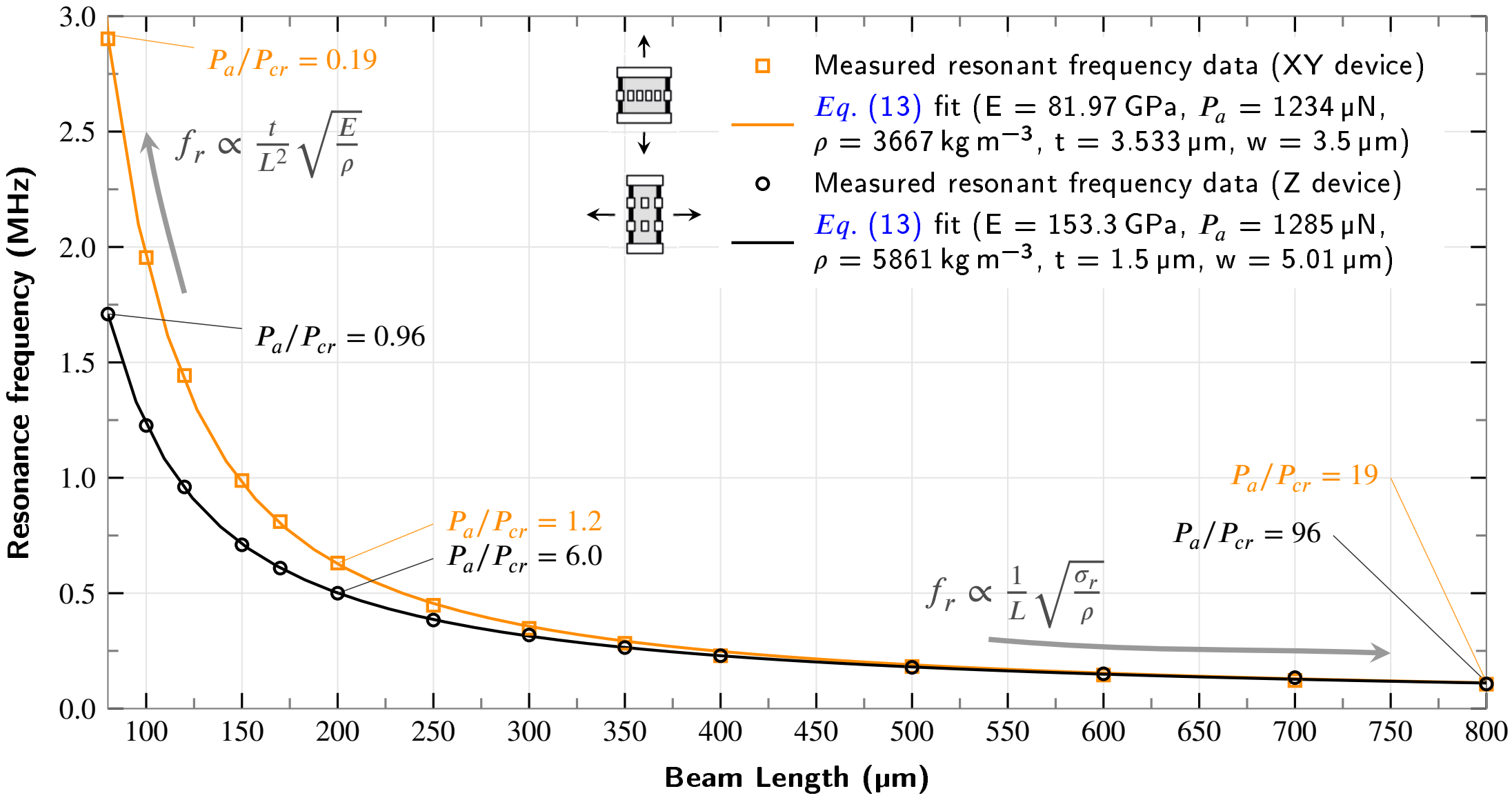

**Figure 12:** Beam to string transition. Measured resonance frequencies of clamped-clamped beams of different lengths, and fitted with Eq. (13), which allows to obtain Young's modulus $E$ and tension or axial load $P_a$.

was fitted using Eq. (13). The fit is very good, which indicates that both the Young's Modulus (E) and the tension ($P_a$) (or residual stress $\sigma_r = P_a/(wt)$) are independent of the beam length.

Optical methods have been used to determine $E$ and $\sigma_r$ in several studies[51–53]. Unfortunately, these methods can only extract the effective values of composite beams applicable to the out-of-plane direction. Fortunately, the presented method also allows determination along the in-plane direction.

Three mechanical behaviors or regions may be distinguished in Fig. 12: one, for short beams or low tensile stress ($|P_a/P_{cr}| < 0.2$), where the resonant frequency ($f_r$) mostly depends on the value of $E$ and it is inversely proportional to the length squared[32]; another one for long beams or high tensile stress ($|P_a/P_{cr}| > 50$), where it mostly depends on $P_a$ or $\sigma_r$ and it is inversely proportional to the length, like for cables and membranes[32]; and a third mixed region ($0.2 < |P_a/P_{cr}| < 50$) in which neither $E$ nor $P_a$ can be neglected. The dependence for the 3 regimes may be readily derived from Eq. (13) by taking the appropriate limits. Note that longer beams converge to the same $f_r$ because their ratio $\sigma_r/\rho$ is very similar, by coincidence. Longer beams were not fabricated due to die size limitation, so the maximum length achievable remains to be studied.

Remarkably, beams as long as 800 µm remained functional, indicating a planarity better than 0.9 µm (vertical gap between beam and electrode). Tensile residual stress was used in order to achieve long structures that are planar and robust against temperature excursions as was put forward in Fig. 3 and confirmed with these measurements.

### *Temperature experiments: $Q(T)$ and $f_r(T)$*

*Maximum working temperature. Closed cavity:* A QFN z device (600 µm-long beams with 0.5 µm gap) was put inside an oven and heated up from 26 °C to 198 °C in steps of 10 °C, approximately. Its resonance curve was permanently monitored. After each temperature increment we waited between 20 min and 34 min until a stable temperature was reached and then the measurement was taken. Finally, the device was removed from the oven and measured again. The obtained resonance frequencies and Q factors are plotted in Fig. 13b. The device survived the test.

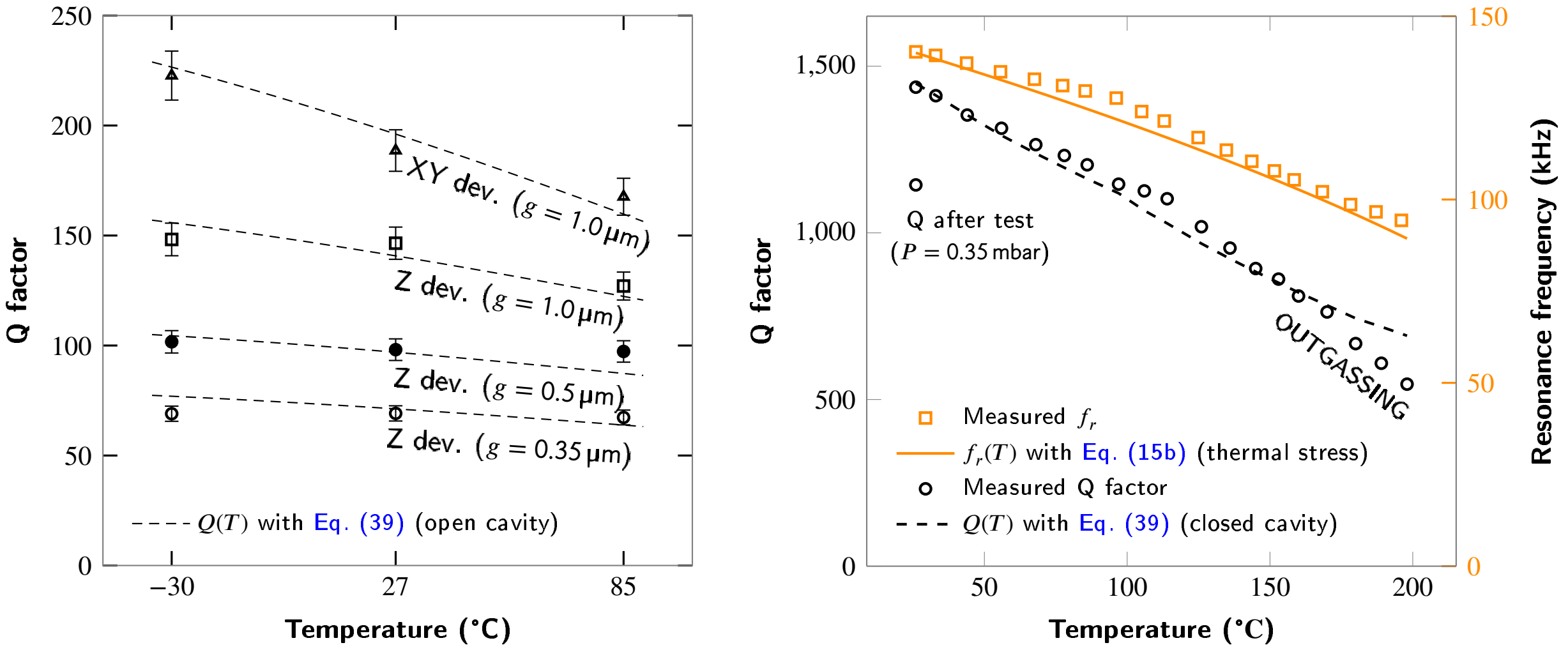

(a) $Q(T)$: Open cavity (z device on wafer at constant $P = 4$ mbar).

(b) $Q(T)$ and $f_r(T)$: Closed cavity (QFN sealed z device at $P = 0.28$ mbar when $T = 26\,°C$ before the test).

**Figure 13:** Measured $f_r$ and Q factor as a function of temperature and predicted values using Eq. (15b) ($f_r$) and Eq. (39) ($Q$).

The resonant frequency decreased as the temperature rose. Eqs. (15b) and (16) predicted well the resonant frequency as it can be appreciated in Fig. 13b. It is important to note that the CTE of the substrate (Silicon) must be included in the equation. Otherwise, the $f_r$ decrease with temperature will be overestimated. Constant values of $E_i\alpha_i$ with respect to temperature can be assumed given that this is approximately true for the BEOL materials of the beams. If not, their temperature dependence ($E_i(T)$ and $\alpha_i(T)$) should be introduced into Eq. (16).

Equation (37), with the values listed in Table 1 for the xy device with $g = 0.5\,\mu m$ and $Q = 1436$, was used to estimate the pressure inside the cavity at the beginning of the experiment: 0.28 mbar. At this pressure, TED ($Q_{TED} > 20000$) is negligible compared to air damping ($Q_{air} \approx 1400$). Therefore, we can rely on Eq. (37), rather than Eq. (38), for Q estimation and its dependence with temperature. In addition, the $f_c$ effect may be neglected in Eq. (37) given that $f_r << f_c$ ($f_r = 140\,\text{kHz}$ and $f_c$ is in the order of MHz). In a closed cavity, the pressure increases linearly with temperature. So, the mean free path $\lambda_g \propto T/P$ remains constant when the temperature changes. Contrarily, in an open cavity at constant pressure the mean free path increases linearly with temperature. By taking into account these considerations, Q factor dependence with temperature may be greatly simplified from Eq. (37) to:

$$Q \propto \frac{f}{\mu_0} \cdot \lambda_g \propto \begin{cases} \dfrac{f}{\mu_0} & \begin{array}{c}\text{Closed Cavity}\\ (\lambda_g \propto T/P = \text{constant})\end{array} \\[2ex] \dfrac{f}{\mu_0} \cdot T & \begin{array}{c}\text{Open Cavity}\\ (\lambda_g \propto T,\ P = \text{constant})\end{array} \end{cases} \tag{39}$$

Note that this approximation is not valid in other situations such as lower pressures where TED dominates, in the fluidic regime or when the resonant frequency is comparable to the cut-off frequency. Full Eq. (38) shall be used in those cases.

Initially, the gas in the cavity is Argon at 1.4 µbar. But during the packaging steps other species are out-gassed, the pressure rises to generally more than 100 µbar and the final gas composition is unknown. In Eq. (39), the viscosity of the surrounding gas $\mu_0$ was calculated using the formulas given in Lemmon and Jacobsen[54] for air. For Nitrogen or using the tabulated values for air[55] the results are very similar. The viscosity of gases increases with temperature and it does not change appreciably with pressure.

Using Eq. (39) the predicted Q factor as a function of the temperature is plotted in Fig. 13b. The measured values follow the curve within 7%, except in the higher temperature region, where the measured Q factor drops appreciably.

This indicates that, above $150 - 175\,°C$ an outgassing mechanism is exacerbated. As a consequence, after the device is cooled down to room temperature, the Q factor decreased from 1436 to 1146, indicating that the pressure inside the cavity increased from 0.28 mbar to 0.35 mbar, approximately. The datapoint at 198 °C can be used to estimate the Q factor once the device returns to room temperature using the closed cavity case of Eq. (39):

$$Q^{26\,°C} = Q^{198\,°C} \frac{f^{(26\,°C)}}{f^{(198\,°C)}} \frac{\mu_0^{(198\,°C)}}{\mu_0^{(26\,°C)}} = 546 \frac{138}{94.3} \frac{25.95}{18.47} = 1123 \approx Q^{26\,°C}_{measured} = 1143 \tag{40}$$

which confirms the validity of Eq. (39) for a closed cavity.

*Maximum working temperature. Open cavity:* In order to confirm the validity of Eqs. (37) and (39) for the open cavity case, four devices were measured at wafer level before sealing at 3 different temperatures and constant pressure (4 mbar). The results are shown in Fig. 13a. The error bars represent the Q uncertainty caused by the measurement noise. The dashed lines predict reasonably well the measured data and were produced using Eq. (37) and a $\gamma$ 10% smaller than in Table 1. Note that the on-wafer measured devices are from a wafer and lot different from the QFN devices, and that 10% is accountable for the expected process variability. In the open cavity case the Q factor variation is significantly smaller (around 2-4 times) than for the QFN devices (closed cavity). The reason is that the $f_r/\mu_0$ variation is partially compensated by the mean free path variation in the open cavity case, as Eq. (39) shows.

## Performance

### *Noise and heading accuracy*

The heading accuracy (Eq. (6)) and the noise floor (Eq. (10)) were plotted in Fig. 14 using data from the measured devices, which provided $Q$, $f_r$, $\eta$ and $L_m$. The continuous curves were calculated using the Q factor from Eq. (38) and the lumped mass $M = \eta^2 L_m$ obtained from the G-B curves of the measured devices. The lowest limiting noise values are around $2 - 3\,\mathrm{nT}/\sqrt{\mathrm{Hz}}$ and the best case heading accuracy around $0.006°/\sqrt{\mathrm{Hz}}$ were achieved for the Z and XY longest devices, and using a Lorentz current of 600 µA. These numbers are state-of-the-art for 3 axis Lorentz-force magnetometers: the lowest noise ($10\,\mathrm{nT}/\sqrt{\mathrm{Hz}}$) had been reported in Kyynarainen et al.[11]. They could be potentially improved with longer beams or higher current which does not neccesarilly mean higher power consumption if the Lorentz wires are designed wider accordingly.

The approximate pressure range of a QFN device is depicted in Fig. 14 as a light gray area. For QFN devices, the lowest noise values are around $7 - 10\,\mathrm{nT}/\sqrt{\mathrm{Hz}}$ with an associated heading accuracy around $0.02 - 0.03°/\sqrt{\mathrm{Hz}}$.

The Lorentz current can be reduced as desired and the noise floor will increase linearly. The dissipated power ($\propto I^2 R$) will decrease quadratically. So the system may be run in a number of different configurations depending on the requirements. For example, if power consumption specs are stringent, $i_L^{peak} = 100\,\mu A$ may be used, and the voltage drop along the MEMS would be less than 0.5 V. In that case, the current may be reused for the electronics. The equations presented in this work allow to calculate the expected performance for other configurations.

System-level simulations at the resonant frequency show that SNR (Signal-to-Noise Ratio) due to the mechanical Brownian noise decreases with Q with a factor of $\sqrt{Q}$ as Eq. (10) shows, while the SNR of the electronics increases linearly with Q, as its noise only depends on the parasitic capacitances and the circuit current consumption. With a low-noise amplifier of 50 µA biasing ($i_{bias}$), the Brownian noise dominates at the pressure level inside the QFN packaged devices. For a given device size and power constraint, system-level power-noise optimization shall redistribute the $i_{bias}/i_L$ current ratio for optimal overall SNR for minimum power consumption.

### *Sensitivity*

The sensitivity to magnetic fields was measured using a Helmholtz coil. The applied magnetic field ranged from −473 µT to 473 µT. A square Lorentz current of 50.14 µA (45.14 $\mu A^{RMS}$ at $f_r$) was injected in phase with the Vac voltage of the impedance analyzer (10 $mV^{RMS}$). Phase alignment was achieved with the system described in Sánchez-Chiva et al.[36].The Lorentz force is added to the electrostatic force and modifies the measured G-B curve (and $R_m$, $L_m$ and $C_m$ values) as described in the MEMS electrical model section. Eight magnetic sweeps were performed so the G-B curve was measured eight times at each of the seven B values to reduce measurement noise and/or temperature drifts. The variance of each group of 8 points was used to perform a weighted linear regression that yielded the $1/R_m$, $1/L_m$ and $C_m$ values ($R^2 > 0.999$) as a function of the magnetic field (see Table 3). It was done for one z and one xy

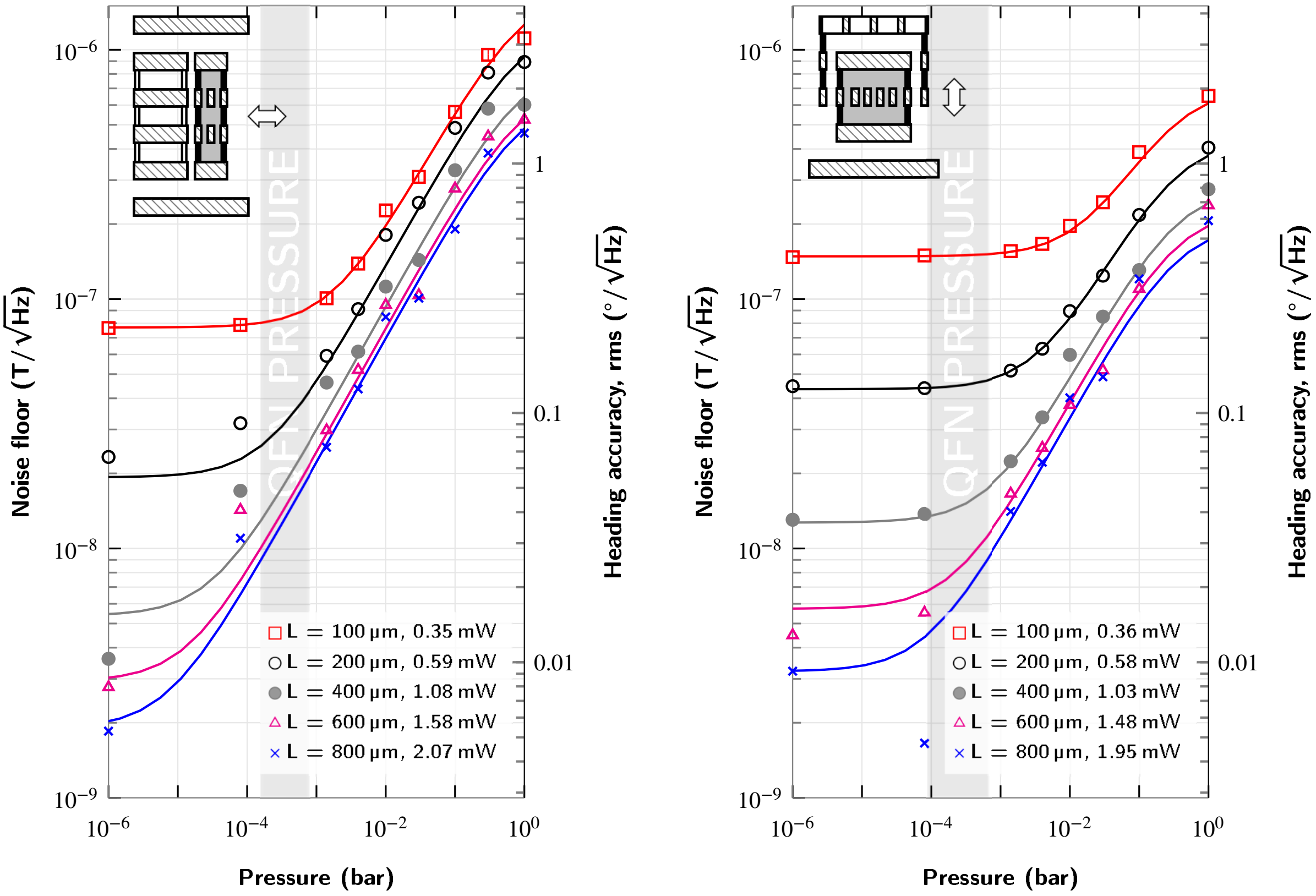

(a) Z device: $g = 0.35\,\mu\text{m}$, $n_B = 10$, $n_w = 2$.

(b) XY device: $g = 0.9\,\mu\text{m}$, $n_B = 6$, $n_w = 4$.

**Figure 14:** MEMS brownian noise floor and heading accuracy versus pressure for different designs. Lorentz current = 0.6 mA (square wave), which respects maximum electromigration current. Noise is inversely proportional to Lorentz current. The continuous lines were obtained using the lumped mass derived from G-B curves and the Q given by Eq. (38), which models well experimental data. The data points were obtained from G-B curves at different pressures. In dark gray, range of pressures of QFN packaged devices. Heading accuracy or angle error is calculated assuming $B_{earth} = 20\,\mu\text{T}$.

device. The $\Omega$ value obtained from the fit is displayed in the table, along with the sensitivity calculated using Eqs. (26) and (27). It is important to note that the Q factor of the chosen z and xy devices is low: the worst case scenario after QFN packaging. The expected sensitivity of a nominal QFN device is between 1.25 and 2.5 times higher than the measured ones for the xy and z devices, respectively.

The measured $\Omega$ is similar to the theoretical value (within 15%). This similarity is an indication that our theoretical estimation of the electrostatic and magnetic forces (and associated Brownian noise) are correct. This 15% discrepancy is perfectly accountable by the inherent process variability, which affects lateral gaps and layer thicknesses. In general, $\Omega$ must be calculated using simulations to estimate the electrostatic force correctly. Fortunately, the electrostatic force of the z device may be calculated analytically quite accurately, with parallel plate assumption. For illustration purposes, its $\Omega$ value may be calculated as follows:

$$\Omega_z = \frac{F_m}{F_e} = \frac{i_L n_w L B}{\epsilon_0 \frac{V_{DC} V_{AC} L w}{g^2}} = 101,5 \cdot B(T) \tag{41}$$

where $w = 5\,\mu\text{m}$ is the width of the electrostatic area of one beam. The parallel plate assumption yields $\Omega_z = 0.048$ for $B = 473\,\mu\text{T}$, similar to the measured and simulated values shown in Table 3.

The maximum conductance value $G_{max}$ may be used to obtain the sensitivity as in Sánchez-Chiva et al.[36]. It is not used given that variability in the $R_p$ values, for example due to electrostatic coupling between sense and Lorentz wire, introduces errors in the measurement that are overcome using the $R_m$, $L_m$ and $C_m$ values.

| Device type | Setup Helmholtz coil | Measurement conditions (RMS) | Fitted linear eq. (Eqs. (22) to (24)) $Y_m = Y_m^{B=0} \cdot (1+\Omega)$ | Magnetic sensitivity (RMS) |
|---|---|---|---|---|
| Z<br>$L = 600\,\mu m$<br>$g = 0.5\,\mu m$<br>$Q = 770$<br>$n_w = 2,\ n_B = 10$ | 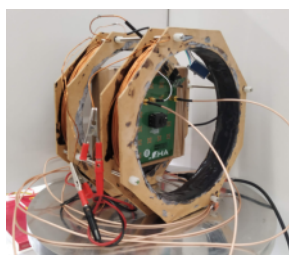 | $V_{AC} = 10\,mV$<br>$V_{DC} = 0.5\,V$<br>$i_L = 45.14\,\mu A$ | $1/R_m = 1.075 \times 10^{-6} \cdot (1 + 96.00 \cdot B(T))$<br>$1/L_m = 1.255 \times 10^{-3} \cdot (1 + 97.77 \cdot B(T))$<br>$C_m = 1.543 \times 10^{-15} \cdot (1 + 97.80 \cdot B(T))$ | $\Omega_z = 96.89 \cdot B(T)$<br>$S_z = 1.05\,\mu A/T$<br>$S'_z = 46.3\,\mu A/(T\,V\,mA)$ |
| XY<br>$L = 600\,\mu m$<br>$g = 0.9\,\mu m$<br>$Q = 1700$<br>$n_w = 4,\ n_B = 6$ | 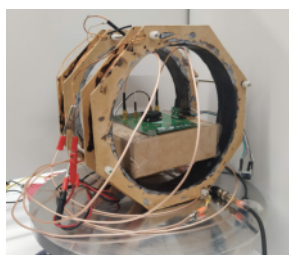 | $V_{AC} = 10\,mV$<br>$V_{DC} = 1\,V$<br>$i_L = 45.14\,\mu A$ | $1/R_m = 0.3958 \times 10^{-6} \cdot (1 + 311.0 \cdot B(T))$<br>$1/L_m = 1.9287 \times 10^{-4} \cdot (1 + 306.8 \cdot B(T))$<br>$C_m = 2.851 \times 10^{-16} \cdot (1 + 306.8 \cdot B(T))$ | $\Omega_{xy} = 308.9 \cdot B(T)$<br>$S_{xy} = 1.22\,\mu A/T$<br>$S'_{xy} = 27.1\,\mu A/(T\,V\,mA)$ |

**Table 3**
Measured $\Omega$ and sensitivity $S$ (Eq. (26)) and $S'$ (Eq. (27)) of two QFN-packaged devices.

Finally, the very high coefficient of determination obtained in non-weighted fits ($R^2 > 0.999$) confirms the high linearity of Lorentz-force magnetometers[56,57], mainly limited by the non-linearity of the motion detection capacitors ($C(x)$). It also proves that this new method for measuring the sensitivity of Lorentz-force magnetometers works well. In fact, the $\Omega$ ratio may be found with very good accuracy (we estimate better than 2% in our case).

### *Offset and shielding efficiency*

One the best features of the presented magnetometers is their Lorentz current shielding from the sense electrode. The theoretical models have been already presented. Now, in order to check the shielding efficiency experimentally, the G-B curve was measured while a Lorentz current of 52 µA (peak value of square wave) in phase with impedance analyzer alternating voltage ($V_{1-2}$ in Fig. 8) was injected into the MEMS. The external magnetic field was compensated with Helmholtz coils. The experiment was repeated but this time using a Lorentz current in anti-phase. The resultant admittance curves are shown in Figs. 15c and 15g for the z device, and in Figs. 15a and 15e for the xy device.

An appreciable constant offset due to the Lorentz current interference effect was measured, especially in the susceptance (B) curves. This corresponds to the electrical interference term in Eq. (3). The interference magnitude ($V_{sh}^{int}$) and the shielding efficiency ($V_{sh}^{int}/V_w$) can be deduced from the G-B curves as follows:

The impedance analyzer extracts the admittance from the nodes 1-2 of Fig. 8 by applying an AC voltage ($V_{1-2}$) and measuring the current flow. The interference caused by a Lorentz current $+i_L$ introduces an additional unwanted current $i^{int}$ flowing through 1-2 and this, in turn, is seen as an admittance change. The admittance offsets ($Y_{off} = G_{off} + jB_{off}$) shown in the figures are produced by a $2 \cdot i_L$ current, so the offset must be divided by a factor of two in order to calculate $i^{int}$:

$$i^{int} = \frac{Y_{off}}{2} \cdot V_{1-2} = \frac{G_{off} + jB_{off}}{2} \cdot V_{1-2} \tag{42}$$

Consequently, an interference voltage $V_{sh}^{int}$ is also added on top of the shield voltage $V_{sh}$. We can calculate this voltage simply as:

$$V_{sh}^{int} = Z_{sh} \cdot i^{int} \tag{43}$$

where $Z_{sh}$ is the output impedance of the impedance analyzer. The coupling capacitance between wire and shield $C_{ws}$ was measured. Then, $V_w$ can also be readily deduced after substituting equations Eqs. (42) and (43) into Eq. (11), which yields:

$$V_w \approx \left(B_{off} - jG_{off}\right) \cdot \frac{V_{1-2}}{2\omega C_{ws}} \tag{44}$$

Note that $V_w$ is an equivalent AC voltage that creates the observed interference. It is just a fraction of the real voltage drop $V^{drop}$ along the Lorentz wire, caused by the fabrication inherent variability resulting in a small lack of symmetry, non-compensated AC voltage and its associated interference, already mentioned in the Materials and methods section. The real voltage drop is readily calculable with the resistance value of the Lorentz wire. Therefore, the interference reduction achieved with symmetric Lorentz routing design can also be estimated as $V_w/V^{drop}$.

| Device type | Measurement conditions | Device impedances | Admittance offset | $V_{drop}$ ($i_L R_{wire}$) | | $V_w$ (RMS) Eq. (44) | | $V_{sh}^{int}$ (RMS) Eq. (43) | Magnetic offset |
|---|---|---|---|---|---|---|---|---|---|
| Z<br>$L = 600\,\mu m$<br>$g = 0.5\,\mu m$<br>$Q = 770$<br>$n_w = 2,\ n_B = 10$ | $i_L = 51.1\,\mu A^{peak}$<br>= square wave<br>$V_{1-2}^{dc} = 0.5\,V$<br>$V_{1-2}^{ac} = 10\,mV^{RMS}$ | $C_{ws} = 6.20\,pF$<br>$C_{ss} = 1.36\,pF$<br>$Z_{sh} = 25\,\Omega$<br>$R_{wire} = 4.50\,k\Omega$ | $G_{off}^Z = 0.459\,\mu S$<br>$B_{off}^Z = 3.39\,\mu S$ | 295 mV | → $\left(\frac{1}{45}\right)^*$ | 3.08 mV | → $\left(\frac{1}{7184}\right)^{**}$ | 0.428 µV | 0.43 µT |
| XY<br>$L = 600\,\mu m$<br>$g = 0.9\,\mu m$<br>$Q = 1575$<br>$n_w = 4,\ n_B = 6$ | $i_L = 52.5\,\mu A^{peak}$<br>= square wave<br>$V_{1-2}^{dc} = 1\,V$<br>$V_{1-2}^{ac} = 10\,mV^{RMS}$ | $C_{ws} = 3.56\,pF$<br>$C_{ss} = 2.9\,pF$<br>$Z_{sh} = 25\,\Omega$<br>$R_{wire} = 4.22\,k\Omega$ | $G_{off}^Z = 0.321\,\mu S$<br>$B_{off}^Z = 2.88\,\mu S$ | 282 mV | → $\left(\frac{1}{26}\right)^*$ | 5.17 mV | → $\left(\frac{1}{14301}\right)^{**}$ | 0.362 µV | 0.13 µT |

**Table 4**
Interference/offset reduction achieved with Lorentz routing symmetric design (*) and Lorentz current shielding (**).

By looking at Eq. (44), if $V_w$ is perfectly in phase or in anti-phase with $V_{1-2}$, then $G_{off}$ should be zero. It turned out to be small, but not zero. From the ratio $B_{off}/G_{off}$ we can deduce that $V_w$ was in 83° phase with $V_{1-2}$ during the measurements, and not perfectly in phase.

The measurement conditions and the amount of electrical interference obtained from Eqs. (11) and (42) to (44) are shown in Table 4. We can see that the shielding electrode reduced the interference voltage by a 4 order magnitude factor, approximately. This value is highly dependent on the impedance connected to the shield electrode. In an ASIC we estimate similar shielding attenuation factors. Additionally, the Lorentz routing design reduced the interference between 25 and 50 times. In total, a 5-6 order of magnitude reduction was achieved.

The interference adds an offset to the admittance curves (electrical interference term in Eq. (2)). This offset may be almost eliminated by using the current chopping technique[12,20] mentioned in the introduction. Unfortunately, this technique cannot eliminate the magnetic offset (electrostatic excitation term in Eq. (2)). Let us apply this technique to the measured data by subtracting the $+i_L$ and $-i_L$ curves, and dividing by two, so the result corresponds to $+i_L$: the resultant curves were plotted in Figs. 15b, 15d, 15f and 15h. A small resonance peak was found. The peak was fitted to the MEMS electrical model and, by using the sensitivity equations from Table 3, we found that it corresponds to a magnetic field of approximately 100 µT. We think this is caused by the permanent magnetization of the prototyping package and/or measuring setup. One reason is that the magnetic field has the same magnitude but opposite direction for the z and xy device: note that the interference AC voltage would excite electrostatically the device and create peaks in the same direction, so is an indication that they correspond to a physical magnetic field. Additionally, the observed peaks (100 µT) are between 2 and 3 orders of magnitude larger than the one created by the interference voltage $V_{sh}^{int}$ ($0.1-0.4\,\mu T$), not observed in these measurements. The magnetic offsets were calculated by setting $\Omega = 1$ (in Eq. (41) for the z device) and obtaining the correspondence between magnetic field and $V_{AC}$.

In conclusion, the current chopping technique in conjunction with the beam shielding successfully eliminated the electrical interference. In addition, the magnetic offset was reduced almost 6 orders of magnitude (4 orders due to shielding and 2 orders due to symmetric Lorentz routing), down to, at most, 0.4 µT, approximately. We believe this is an important achievement given that this magnetic offset cannot be compensated with the current chopping technique.

## Yield and reliability tests

### *Yield after QFN packaging*

The two types of z and xy devices shown in Figs. 6a and 7 and in Table 3 were packaged into QFN packages. The chosen variants were formed by 600 µm-long c-c beams. The xy and z devices occupy and area of $680x185\,\mu m^2$ and $690x210\,\mu m^2$, respectively. They were measured and the results and expected cavity pressure, noise floor and associated best heading accuracy are summarized in Table 5. The cavity pressure was deduced from the Q factor and resonant frequency using Eq. (38) and values from Table 1. Noise floor and heading accuracy can be found in Fig. 14. Results showed that the z-yield is high but the xy-yield was significantly lower. We will show in the following section

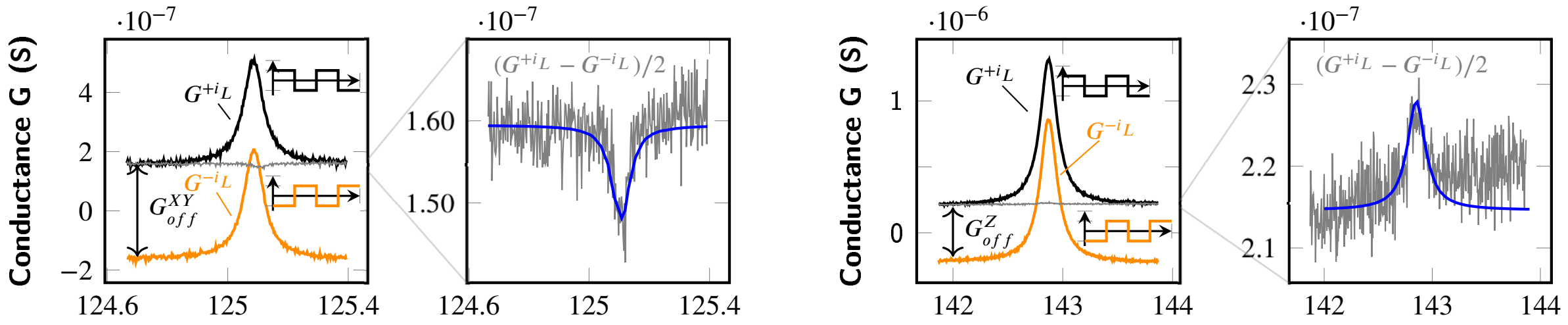

(a) Offset in conductance (G) due to Lorentz current (XY Device).

(b) Magnetic resonance related to $+i_L$ and fit. Conductance (G). XY.

(c) Offset in conductance (G) due to Lorentz current (Z Device).

(d) Magnetic resonance related to $+i_L$ and fit. Conductance (G). Z.

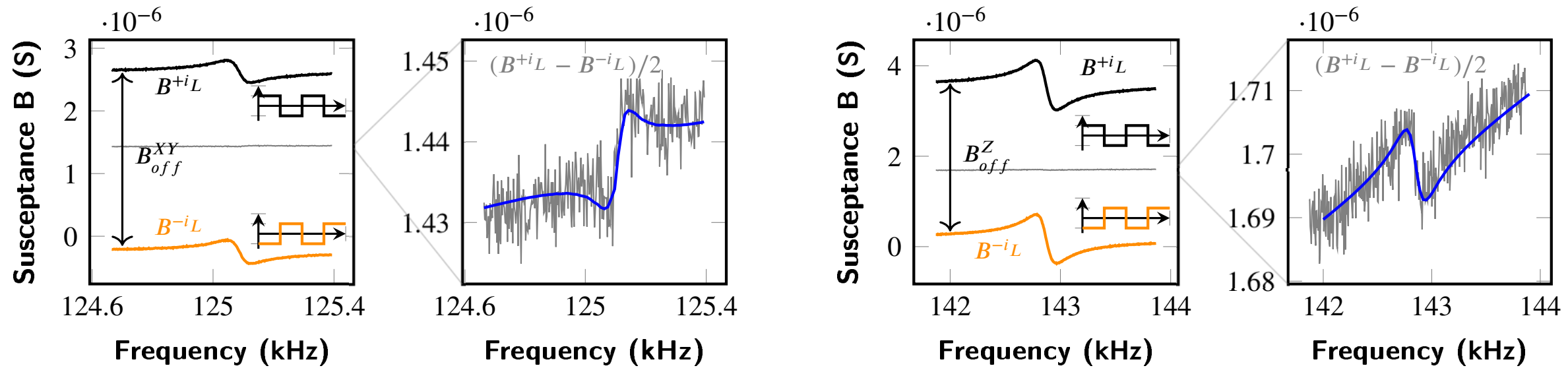

(e) Offset in susceptance (B) due to Lorentz current (XY Device).

(f) Magnetic resonance related to $+i_L$ and fit. Susceptance (B). XY.

(g) Offset in susceptance (B) due to Lorentz current (Z Device).

(h) Magnetic resonance related to $+i_L$ and fit. Susceptance (B). Z.

**Figure 15:** Offsets in the G-B curves due to positive (black) and negative (orange) Lorentz current interference. Their difference shows a resonance (gray).

| Dev. | $f_{r\,-\text{sample min}}^{\,+\text{sample max}}$ | $Q_{-\text{sample min}}^{+\text{sample max}}$ | $P_{-\text{sample min}}^{+\text{sample max}}$ | Noise floor (RMS) | Heading accuracy (RMS) | Yield |
|---|---|---|---|---|---|---|
| Z | $144^{+15}_{-6}$ kHz | $1687^{+1122}_{-1093}$ | $270^{+520}_{-114}$ μbar | $15^{+5}_{-5}$ nT/$\sqrt{\text{Hz}}$ | $0.045^{+0.015}_{-0.015}$ °/$\sqrt{\text{Hz}}$ | 118/124 (95%) |
| XY | $133^{+9}_{-10}$ kHz | $2049^{+935}_{-966}$ | $245^{+245}_{-157}$ μbar | $9.5^{+2.5}_{-2.5}$ nT/$\sqrt{\text{Hz}}$ | $0.027^{+0.009}_{-0.009}$ °/$\sqrt{\text{Hz}}$ | 68/114 (53%) |

**Table 5**
Variability and yield of QFN devices.

how the xy device was redesigned to improve this yield. The main failure mode was a too high G value, indicating a short-circuit between sense and shield electrode. Statistical data will show that this may be caused by a non-optimum sense electrode or coupling link design.

### *Temperature robustness and yield improvement*

On-wafer devices were subjected to one of the two stringent thermal profiles shown in Fig. 16, which are close to altering the CMOS electronics performance[58–60]. Also, Aluminum suffers a significant Young's modulus softening at those temperatures[61]. Withstanding high temperatures for several minutes is very important so the device may be subjected to outgassing thermal treatments (annealings)[62,63] and other post processes. In our experience, a MEMS device that shows high yield after these tests also shows high yield after packaging into QFN, and viceversa. Therefore, it can be used as a quick and cheap method to foresee yield issues before QFN packaging.

The devices were measured before and after the experiment and the yield results are summarized in Tables 6 to 8. The number of mechanical couplings and length of the sensing structures (sensing plate length x number of repetitions) is shown for each device type and length.

Results show that, with proper design, very robust CMOS-MEMS structures may be manufactured. For example, the length of the sensing electrodes of the xy devices had an important effect on the robustness against high temperature. In fact, older devices (not shown here) with longer sensing plates showed much lower yield. Shortening the sense

(2021)

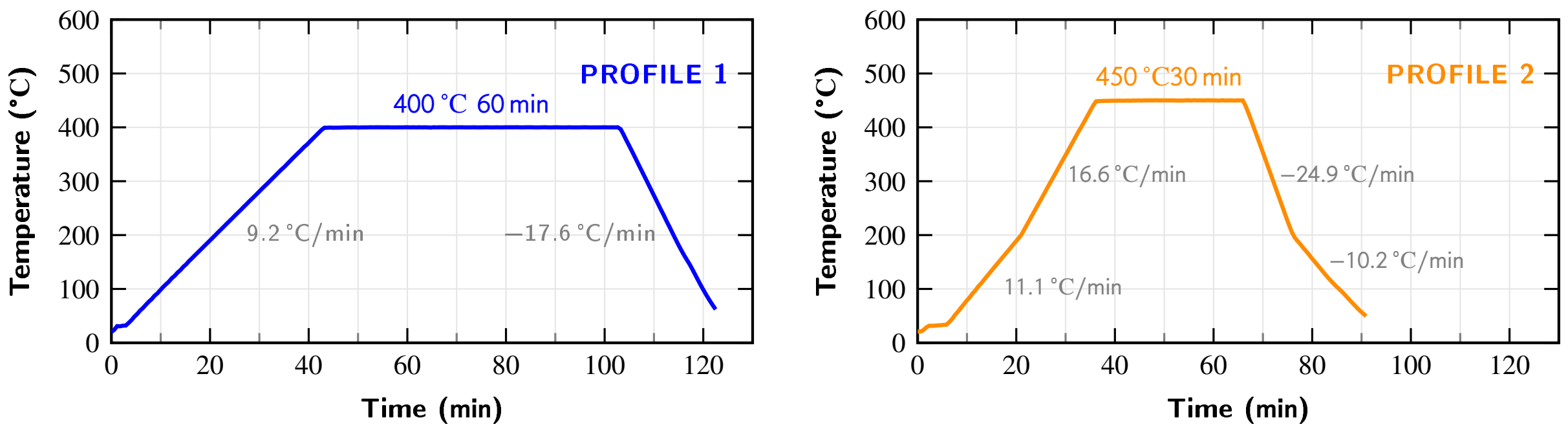

**Figure 16:** Profiles applied to test the device robustness against temperatures that are close to altering the CMOS electronics performance.

| XY device with long sensing plates | | | | | |
|---|---|---|---|---|---|
| Dev. L (µm)<br>Sense L (µm)<br>Couplings | 300 | 400<br>100x4<br>4 | 500<br>82x6<br>6 | 600<br>74x8<br>6 | TOTAL |
| Yield before 400 °C 1 h | | 13/13 (100%) | 11/13 (85%) | 13/13 (100%) | 37/39 (95%) |
| Yield after 400 °C 1 h | | 7/13 (54%) | 5/13 (38%) | 10/13 (77%) | 22/39 (56%) |
| Yield before 450 °C 0.5 h | | 13/13 (100%) | 13/13 (100%) | 12/13 (92%) | 38/39 (97%) |
| Yield after 450 °C 0.5 h | | 5/13 (38%) | 2/13 (15%) | 0/13 (0%) | 7/39 (18%) |

| XY device improved with shorter sensing plates (Fig. 6a) | | | | | |
|---|---|---|---|---|---|
| Dev. L (µm)<br>Sense L (µm)<br>Couplings | 300 | 400<br>33x6<br>2 | 500 | 600<br>36x7<br>6 | TOTAL |
| Yield before 400 °C 1 h | | 12/13 (92%) | | 13/13 (100%) | 25/26 (96%) |
| Yield after 400 °C 1 h | | 12/13 (92%) | | 13/13 (100%) | 25/26 (96%) |
| Yield before 450 °C 0.5 h | | 12/13 (92%) | | 13/13 (100%) | 25/26 (96%) |
| Yield after 450 °C 0.5 h | | 5/13 (38%) | | 3/13 (23%) | 8/26 (31%) |

**Table 6**
Yield results of the two high temperature tests on xy devices.

| XY-4M device with sensing fingers (Fig. 6b), several couplings | | | | | |
|---|---|---|---|---|---|
| Dev. L (µm)<br>Sense L (µm)<br>Couplings | 300<br>28x4<br>2 | 400<br>28x6<br>4 | 500<br>28x8<br>4 | 600<br>28x10<br>6 | TOTAL |
| Yield before 400 °C 1 h | 13/13 (100%) | 13/13 (100%) | 13/13 (100%) | 13/13 (100%) | 52/52 (100%) |
| Yield after 400 °C 1 h | 13/13 (100%) | 10/13 (100%) | 5/13 (100%) | 3/13 (100%) | 52/52 (100%) |
| Yield before 450 °C 0.5 h | 13/13 (100%) | 13/13 (100%) | 13/13 (100%) | 13/13 (100%) | 52/52 (100%) |
| Yield after 450 °C 0.5 h | 13/13 (100%) | 10/13 (77%) | 5/13 (38%) | 3/13 (23%) | 31/52 (60%) |

| XY-4M device with sensing fingers (Fig. 6b), 1 coupling | | | | | |
|---|---|---|---|---|---|
| Dev. L (µm)<br>Sense L (µm)<br>Couplings | 300<br>28x6<br>1 | 400<br>28x6<br>1 | 500<br>28x6<br>1 | 600<br>28x8<br>1 | TOTAL |
| Yield before 400 °C 1 h | 13/13 (100%) | 13/13 (100%) | 13/13 (100%) | 13/13 (100%) | 52/52 (100%) |
| Yield after 400 °C 1 h | 13/13 (100%) | 13/13 (100%) | 13/13 (100%) | 13/13 (100%) | 52/52 (100%) |
| Yield before 450 °C 0.5 h | 13/13 (100%) | 13/13 (100%) | 13/13 (100%) | 13/13 (100%) | 52/52 (100%) |
| Yield after 450 °C 0.5 h | 13/13 (100%) | 13/13 (100%) | 13/13 (100%) | 11/13 (85%) | 50/52 (96%) |

**Table 7**
Yield results of the two high temperature tests on xy-4m devices.

electrode length allowed to improve the yield of the 3-metal xy device from 56% to 96% in the 400 °C-1 h test. Also, it was found that the number of couplings should be minimized: it increased the yield of the xy-4m device with sensing fingers from 60% to 96% in the 450 °C-0.5 h test, achieving similar yields to the z device. This is strong evidence that the yield after QFN packaging of the xy-4m device would be similar to that of the z device.

Implementing a large number of design variants proved a decisive factor in the yield optimization process, which also allowed to reduce the number of iterations and therefore development time.

(2021)

| Z device with gap = 0.5 µm (Fig. 7), 1 coupling | | | | | |
|---|---|---|---|---|---|
| Dev. L (µm)<br>Sense L (µm)<br>Couplings | 300<br>37x8<br>1 | 400<br>50x8<br>1 | 500<br>40x12<br>1 | 600<br>37x16<br>1 | TOTAL |
| Yield before 400 °C 1 h | 13/13 | 13/13 (100%) | 13/13 (100%) | 13/13 (100%) | 52/52 (100%) |
| Yield after 400 °C 1 h | 13/13 (100%) | 13/13 (100%) | 12/13 (92%) | 12/13 (92%) | 50/52 (96%) |
| Yield before 450 °C 0.5 h | 13/13 (100%) | 13/13 (100%) | 13/13 (100%) | 13/13 (100%) | 52/52 (100%) |
| Yield after 450 °C 0.5 h | 13/13 (100%) | 13/13 (100%) | 13/13 (100%) | 13/13 (100%) | 52/52 (100%) |

**Table 8**
Yield results of the two high temperature tests on z devices.

### *Magnetic robustness*

In order to test the robustness against large magnetic fields the z device G-B curve was measured while using a square wave Lorentz current of 50 µA in the presence of a magnetic field of variable intensity. The maximum applied magnetic field was estimated in around 32 mT. It was created with a magnet placed on top of the QFN and a Helmholtz coil, generating 24 mT and 8 mT each, respectively. No malfunction was found during or after the test. Due to the magnetization of external components, offsets of around a few hundreds of µT remained when the magnetic field was set to zero. In addition, non-linear behavior became slightly apparent when vibration amplitudes reached around 20 – 30% of the gap which took place when the magnetic field was around 14 – 20 mT.

### *Shock tests*

The QFN z-device underwent several shock tests along the 3 directions. The impacts were performed manually. A commercial triaxial piezoelectric accelerometer (834M1-6000)[64] was used as the reference accelerometer. Its maximum acceleration range is ±6000 g, so that was the maximum shock that could be recorded. An Arduino microcontroller board was also attached to the test assembly, which recorded the 3 axis readings when a given threshold was surpassed and later sent to a PC via serial communication. The accelerometer bandwidth is above 6 kHz with analog output and its sensitivity 5 g/LSB. Its offset was compensated in the Arduino code. Three readings, one for each axis, were taken every 4.5 µs and the total recording time was 9 ms. The QFN, accelerometer and Arduino were assembled together as shown in Fig. 17d.

The devices survived all the tests performed. The recorded accelerations of the 3 strongest shocks are shown in the graphs in Fig. 17. Along the y and z axes they reached 6000 g. The correct functioning after the test was confirmed by measuring the G-B resonance with an impedance analyzer.

In addition, 5 QFN z devices where subjected to two standard mechanical shock tests:

- Method 2002.5, Condition B: MIL-STD-883 1500g 0.5 ms Half Sine, 5 shocks in each direction of 3 mutually perpendicular axes. (30 total)
- Method 2007.3, Condition A: MIL-STD-883 1.5 mm pk-pk / 20g pk min, 20-2000 Hz, 4 sweeps in each of 3 mutually perpendicular axes at 3 octaves/min.

The 5 devices survived the tests.

Finally, one QFN z device underwent a 3 s free fall from a 5.60 m height and landed on a concrete floor at an estimated velocity of 7 km s$^{-1}$. The acceleration during the impact was not recorded. No appreciable damages were observed in the QFN package and the device continued functioning correctly.

## Performance comparison with other magnetometers

A list of the commercial magnetometers used in smartphones may be found in Matyunin et al.[65]. Some of them are compared with 3 axis MEMS magnetometers in Table 9. The recently commercialized TMR technology displays the best noise figure-of-merit (FOM) of commercial magnetometers. However, the Lorentz-force magnetometers built

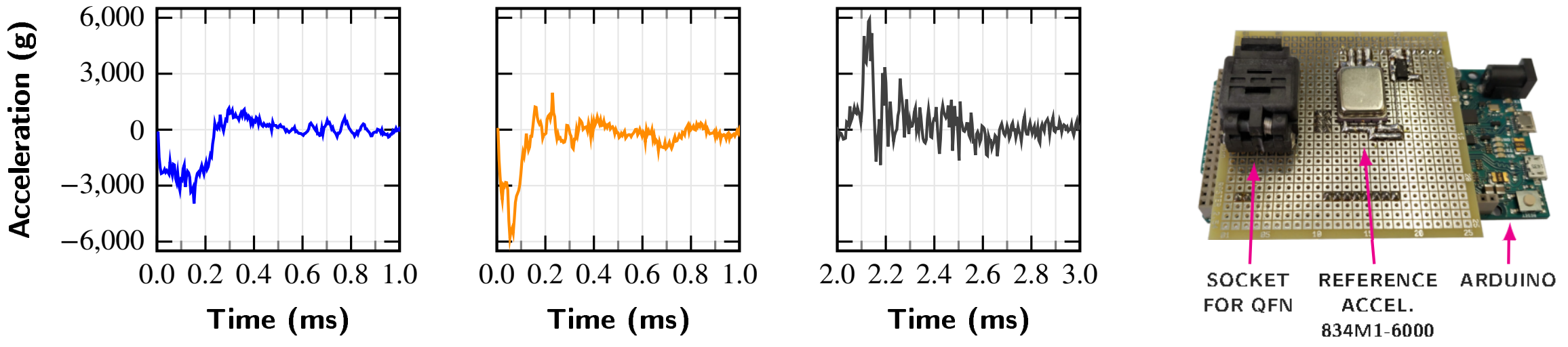

(a) X-shock test results. (b) Y-shock test results. (c) Z-shock test results. (d) Setup for recording acceleration during the shock tests.

**Figure 17:** Recorded acceleration values for the X, Y and Z shock tests (a,b,c) and measurement setup (d).

| | 3D Magnetometer | Technology | Full scale range (±mT) | Current per axis ($\mu A_{rms}$) | Figure of merit* ($\mu T \mu A_{rms}/\sqrt{Hz}$) | Footprint 3 axes ($mm^2$) | Offset ($\mu T$) |
|---|---|---|---|---|---|---|---|
| COMMERCIAL | STMicroelectronics LIS3MDL[66] | AMR | 1.2 | 90 | ~ 10 | 4 | 100 |
| | STMicroelectronics LSM303AGR[67] | AMR | 4.9 | 33 – 316[a] | 6 – 20[a] | 4[b] | 6[c] |
| | Freescale MAG3110[68] | TMR | 1.0 | 2.9 – 300[a] | 3.6 – 19[a] | 4 | 100 |
| | AKM AK8975[69] | Hall | 1.2 | 117 | ~ 100 | 4 | 300 |
| | AKM AK09940[70] | TMR | 1.2 | 40 – 267 | 0.20 – 0.76 | 2.56 | No data |
| | Honeywell HMC5883[71] | AMR | 0.8 | 33/640 | 3.4 | 9 | No data |
| | Bosch BMM150[72] | AMR+Hall | 1.3 | 57 – 1630[a] | 35.6 – 155[a] | 2.43 | 40[b] |
| R&D | Kyynarainen et al.[11] | MEMS LFM | 0.2 | 100 | X/Y: 1.0, Z: 7.0 | > 11.5 | 25 |
| | Laghi et al.[19] | MEMS LFM | 5.5 | 33[d] | X/Y: 6.1, Z: 6.7[f] | 4[d] | 5000[b] |
| | Marra et al.[29],[73] | MEMS LFM | X: 6.0, Y: 5.5, Z: 7.0 | 70[d] | X: 8.4, Y: 5.2, Z: 7.7[f] | 0.53[d] | No data[e] |
| | This work (QFN) | CMOS-MEMS LFM | Z: 32[g] | 0 – 600[d] | X/Y: 1.8, Z: 3.0[f] | 0.4[d] | X/Y: 0.13, Z: 0.43 |

* Normalized for a single axis. For R&D, X, Y and Z axis values are given.
[a] Value varies depending on the selected current/bandwidth.
[b] 3D magnetometer and 3D accelerometer.
[c] Can be reduced to a few μT with manual DC compensation or calibration.
[d] ASIC current consumption/area not included.
[e] Expected to be in the same order of magnitude as Laghi et al.[19] given the design and manufacturing process similarities.
[f] Assuming current is reused for the 3 axes. Otherwise the triple value must be taken.
[g] Tested with a current of 50 μA.

**Table 9**
Comparison of commercial and in research-state 3 axis magnetometers.

with the presented CMOS-MEMS process show very competitive results in QFN packages, compared to both MEMS and commercial devices. For example, the sensor area is the smallest found in 3 axis MEMS magnetometers.

# Conclusions

Integration of MEMS and CMOS is a long-sought objective that would provide significant size, cost and power advantages. However, successful integration has proven to be difficult. In this work, the fabrication process and the design techniques to overcome the main challenges to build reliable CMOS-MEMS devices have been presented.

Three-axis Lorentz-force magnetometers (LFM) were designed, fabricated and extensively characterized: equations that accurately predict the Q factor and resonant frequency of multilayered clamped-clamped beams as a function of temperature, design parameters, and gas pressure from 1 bar to 1 μbar were derived and verified experimentally. TED was the main damping mechanism at low pressures as finite element simulations confirmed. Gas viscosity explained Q factor temperature variations in air damping. Thermal stress accounted for the variation of resonance frequency with temperature. The beam-to-string transition of clamped-clamped beams with the same axial stress but different length was measured and fitted accurately the expected behavior. This demonstrates that accurate modelization of complex multilayered structures built with the BEOL of CMOS is feasible.

Lorentz-force magnetometers do not have magnetic materials, which provides several advantages over other magnetometer technologies. Unfortunately, offsets in LFM are, probably, their main drawback. In this work, the current chopping technique in conjunction with the beam shielding successfully eliminated the electrical interference. In ad-dition, the electrostatic interference/offset, which cannot be compensated with the current chopping technique, was reduced almost 6 orders of magnitude (4 orders due to shielding and 2 orders due to symmetric Lorentz routing) down to 0.13 µT and 0.43 µT for the xy and z axes, respectively.

Despite CMOS technology not being a MEMS process, Brownian noise in the final CMOS-MEMS QFN-packaged devices was between $9.5 - 15\,\mathrm{nT}/\sqrt{\mathrm{Hz}}$ when using a current of 600 µA. A heading accuracy as low as $0.045°/\sqrt{\mathrm{Hz}}$, approximately, may be achieved by a compass formed by the packaged magnetometers. This is similar or better than what commercial magnetometers and state-of-the-art three-axis LFMs built with MEMS-dedicated processes can provide. Apart from the offset and noise benefits, the sensor area is the smallest found in 3 axis MEMS magnetometers. One of the tested devices on wafer reached a Q factor of around 40 000 at 107 kHz. This is equivalent to a Brownian noise level of $2\,\mathrm{nT}/\sqrt{\mathrm{Hz}}$ with a Lorentz current of 600 µA. This is lower than the three-axis LFMs found in the literature. A lower noise level could be achieved with longer beams not fabricated in this work, or higher Lorentz current.

Yield is usually one the major concern in MEMS products. Conveniently, we showed that the final yield of a QFN packaged CMOS-MEMS device can be around 95%. In addition, some device variants withstood very high temperatures with none or little yield loss: 450 °C for 30 min and 400° for 1 h. As summary, robust CMOS-MEMS devices with potential to equal or out-best commercial products is possible using the appropriate design techniques.

# Acknowledgments

The authors would like to thank Laura Barrachina and Sandra Aguilar from Baolab Microsystems for assistance during the measurements. This work was supported by Baolab Microsystems and by the Spanish Ministry of Science, Innovation and Universities, the State Research Agency (AEI), and the European Social Fund (ESF) under project RTI2018-099766-B-I00.

# Author information

## Contributions

J.V conceived the design techniques, designed the sensors, performed simulations, proposed and analyzed the measurements, established the new theoretical models and wrote the manuscript. J.M.S-C and J.V. performed measurements. J.M.S-C and D.F. conceptualized test setups. D.F. and J.M. supervised the research. All the authors discussed the results and revised the manuscript.

## Corresponding author

Correspondence to Juan Valle.

# Ethics declarations

## Conflict of interest

The authors declare that they have no conflict of interest.

(2021)